\documentclass[traditabstract]{aa}  
\pdfoutput=1
\usepackage{graphicx}
\usepackage{amsmath}
\usepackage{txfonts}
\usepackage[]{natbib} 
\usepackage[]{natbib} 
\usepackage[usenames]{color}
\usepackage{xcolor}
\newcommand{\hi} {H{\sc i}}

\newcommand{\mi}{$\mu$m}

\newcommand{\mm}{M\,33}
\newcommand{\Ha}{H$\alpha$}

\begin{document}

\title{Rise and fall of molecular clouds across the M33 disk}

\titlerunning{Molecular clouds across M33 }  
 
\author{Edvige Corbelli
       \inst{1}
        \and 
           Jonathan Braine
       \inst{2}
        \and
           Carlo Giovanardi
       \inst{1}
                 }

   \institute{INAF-Osservatorio Astrofisico di Arcetri, Largo E. Fermi, 5,
             50125 Firenze, Italy\\
             \email{edvige@arcetri.astro.it} 
             \and 
             Laboratoire d'Astrophysique de Bordeaux, Univ. Bordeaux, CNRS, B18N, 
                   all\'ee Geoffroy Saint-Hilaire, 33615 Pessac, France\\
                }

   \date{Received .....; accepted ....}

 \abstract
   
 \abstract
      {We carried out deep searches for CO line emission  in the outer disk of M33, at R$>7$~kpc, and examined the dynamical conditions that can explain variations in the mass distribution of the molecular
      cloud throughout the disk of M33. 
       We used the  IRAM-30~m telescope to search for CO  lines in the outer disk toward 12 faint mid-infrared (MIR) 
       selected  sources  and  in an area of the southern  outer disk hosting MA1, a bright HII region.  
       We detect narrow CO lines at the location of  two MIR sources at galactocentric distances of about 8~kpc that are associated with low-mass young stellar clusters,  and at four locations
       in the proximity of MA1. The paucity of  CO lines at the location of weak MIR-selected sources probably arises because most of them are not star-forming sites in M33,
       but  background sources.  Although very uncertain, the  total molecular mass of the detected  clouds around MA1 is lower than  expected   given the stellar mass of the cluster,
       because dispersal of the  molecular gas  is taking place as  the  HII region expands.
       The mean mass of  the giant molecular clouds (GMCs) in M33 decreases radially by a factor 2 from the center out to 4~kpc, then it stays constant  until it drops at R$>7$~kpc. We suggest that GMCs 
       become more massive toward the center  because of the fast rotation of the  disk, which  drives mass growth by coalescence of smaller condensations  as they cross the  arms.  
       The  analysis of  both HI and CO spectral data gives the consistent result  that corotation of the two main arms in this galaxy is at a radius of  4.7$\pm$0.3~kpc, and spiral shock waves become subsonic beyond 3.9~kpc. 
       Perturbations are quenched beyond 6.5~kpc, where CO  lines have been detected   only around sporadic condensations associated with UV and MIR emission. 
         }
     
   \maketitle   
 
\section{Introduction}

 Our knowledge of molecular clouds and the processes in  the interstellar medium (ISM) that favor the birth of stars is  mostly based on Galactic studies.  The increase in resolution and 
sensitivity has enabled recent extragalactic surveys to study the formation of stars as galaxies evolve and as a function of   galaxy mass, morphology, and  metallicity.  
Nearby galaxies are an ideal place where a global picture of the disk is complemented by detailed observation of the ISM and star-forming  (hereafter SF)  sites. 
In this context, particular attention has been given to  low-luminosity Local Group galaxies with subsolar metallicity: M33 in the north and the Large Magellanic Cloud (LMC) 
in the south. The spiral galaxy M33 is particularly interesting because it is a blue SF galaxy with no evident sign of ongoing or past interactions, as shown by
the most recent proper motion measurements and past orbital history \citep {2017MNRAS.464.3825P}. 
It is a relatively unperturbed spiral with   no bulge \citep{2007ApJ...669..315C} and an extended  warped outer disk \citep{2014A&A...572A..23C}.
This makes it an ideal laboratory to study how the gas settles in the disk and develops instabilities that condense and form stars and/or to determine the role of feedback in triggering
a new generation of stars \citep{2018MNRAS.478.3793D,2018A&A...617A.125C}.    

The  investigation of  individual SF  complexes in  nearby galaxies and  all-disk surveys at infrared and millimeter wavelengths 
(through facilities such as {\it Spitzer, Herschel}, IRAM, and ALMA)  reveals how the ISM and the star formation process differ in chemically young environments 
as compared to more evolved  spirals.   The molecular cloud
 mass spectrum steepens, the conversion of molecular gas into stars is faster, and the CO-to-H$_2$ ratio decreases in lower metallicity environments
  \citep{2012A&A...542A.108G,2005PASP..117.1403R,2008ApJS..178...56F,2009A&A...493..453V,2007ApJS..173..185G,2010A&A...512A..68G,2011MNRAS.415.3439D}.  
M33 is an ideal object to study these differences because its metallicity is only a factor two below solar \citep{2010A&A...512A..63M}
and it retains a clear  disk spiral morphology such that morphological and 
 chemical differences are not mixed. Because of the limited differences, we can continue to use the same tracers as in large spirals.  M33 represents a sort of stepping stone toward
 smaller objects where chemical differences are more extreme and mixed with morphological changes. 
 
Recently, \citet{2017A&A...601A.146C} have estimated the duration of the life cycle of giant molecular clouds (hereafter GMCs) in M33 using the all-disk 
CO J=2-1 survey \citep{2014A&A...567A.118D} to identify 566 GMCs in the SF disk, and a mid-infrared (hereafter MIR) selected sample of 630 young stellar cluster candidates 
(hereafter YSCCs) from the catalog of \citet{2011A&A...534A..96S}.
The GMCs spend 4 Myr in the inactive phase and 10~Myr in the active SF phase.  Sources are in the embedded phase (detected only in the
infrared)  during only 2 Myr of the active phase, and for the remaining time  YSCCs  break  through  the cloud and become also detectable in H$\alpha$ or UV.
The correlation between  GMCs and MIR-selected YSCCs is remarkable: in the active phase,  all GMCs are associated with MIR sources.

There is,  however,  a non-negligible fraction of MIR sources that are not hosted by  GMCs, and this fraction increases
moving radially outward. One possibility is that these MIR sources are associated with less massive molecular clouds below the detection limit, which form low-mass clusters
that are often undetectable in  H$\alpha$ because of the lack of massive stars. Through  the MIR source catalog and CO pointed observations, 
\citet{2011A&A...528A.116C}  have found   low-mass SF complexes associated with molecular clouds    that are weak  in CO J=1-0 and J=2-1 line emission. 
A key question related to this issue is to understand variations in the molecular mass spectrum  beyond galactocentric radii of  about 4~kpc in M33
\citep{2018A&A...612A..51B}. We know that the mass spectrum of molecular clouds changes across the disk of a galaxy \citep{2005PASP..117.1403R,2012A&A...542A.108G,2015ARA&A..53..583H},
but the reason for this, and whether other properties change, is still an open question 
\citep{2010ApJ...725.1159B,2014ApJ...784....3C,2017MNRAS.468.1769F,2017ApJ...836..175K}. The onset of instabilities and  
the ability of the gas to cool and fragment may be not efficient enough to create massive complexes at large galactocentric radii, or it might just be that the spiral pattern cannot 
accumulate  gas and merge clouds to make  more massive complexes beyond corotation, although disk instabilities can still trigger the formation of  filaments 
\citep{1995ApJ...445..591E,2018MNRAS.478.3793D}.  

Outer disks are key places to understand galaxy evolution: 
in the local Universe, $\Lambda$CDM models predict that  large spirals like M31 do not accrete much gas into the disk, while low-mass galaxies,  such as M33,
experience an inside-out growth and feed star formation  through cold gas streams from the intergalactic medium that settle into the  disk \citep{2007A&A...470..843M,
2009ApJ...695L..15W,2009MNRAS.395..160K,2013MNRAS.435..999D,2018MNRAS.479..319F}.  It is therefore
important to understand how pristine outer gas mixes with the galaxy ISM, and in particular, if and how  star formation and metal enrichment  occur beyond
the bright SF disk and if this occurs continuously or in a burst.
Imaging in V and I band with the Subaru/Suprime-Cam  of a  few fields of M33 in the  outer regions  \citep{2011A&A...533A..91G} has revealed  a pervasive diffuse evolved stellar population 
($>$1Gyr), but also a population of younger stars  (100-200 Myr)  where   \hi\ gas overdensities are present in the outer disk.  
An interesting output of MIR  source selection  in the M33 area is that low-luminosity sources are also present in  correspondence with the outer disk
beyond the   SF edge   where the  H$\alpha$ surface brightness drops    \citep{ 2009A&A...493..453V,2011A&A...534A..96S}.  
One possibility is that a background population of faint MIR sources is mixed in projection with truly M33 SF sites and that this becomes the dominant population at large radii.
It may also be, however, that some of these faint MIR sources are small SF sites in the outer disk of M33. In this paper we  use 
sensitive IRAM-30m observations to understand the nature of these faint MIR sources.  
It is conceivable  to detect CO emission in molecular clouds just beyond the SF edge in M33 because the metallicity  is about one-third solar  
 \citep{2011A&A...533A..91G,2010A&A...512A..63M} and the UV field is weak \citep{2005ApJ...619L..67T}. 
It is still unclear which are the conditions  that favor the growth of molecular clouds beyond the SF edge of a galaxy, 
and whether it is possible to find them in the extreme outer disk \citep{1994ApJ...422...92D, 2002ApJ...578..229S, 2007ApJ...669L..73B}. 

In Section~2  we summarize some properties of GMCs and MIR sources across the M33 disk.
In Section~3  we describe  CO observations in the outer disk for a selected MIR sample  and around 
a bright HII region. In Section~4 we discuss some implications of the detected and undetected CO lines, 
such as  the molecular mass of condensations in the outer disk and the nature of faint MIR sources. In Section~5  we show results relative to disk dynamics   
that can be linked to  the observed radial distribution of molecular clouds. Section~6 concludes after we summarize the main results.

\section{Distribution of GMCs and MIR sources across the disk of M33}

M33 has  a gaseous disk with holes and dense filaments related to optically visible flocculent spiral arms. Two of these  have more prominent and numerous HII regions, one is in the northern approaching side  
and the other is in the southern receding side. 
We examine three distinct radial ranges, R$<4$~kpc, $4<R< 7$~kpc, and $R\ge 7$~kpc,  and we refer to these as inner disk  (dominated by the two main arms), the 
 intermediate disk,  and the  outer disk.  In this paper we refer to the SF disk of M33 to indicate the area within galactocentric distances of 7~kpc (inner + intermediate disk). 
The  outer disk  encloses the  extreme regions of the optical disk close to  R$_{25}$ and the warp, and it is beyond the H$\alpha$ luminosity drop  \citep{1989ApJ...344..685K,2009A&A...493..453V}. 
In this section we review the changes in the mean mass of GMCs  
and the 24~$\mu$m flux of MIR  sources  across the M33 disk. We also describe the selection of  the MIR sample for deep CO observations in the outer disk.  

The GMC catalog comprises 566 clouds with masses between 2$\times 10^4$ and 2$\times 10^6$ M$_\odot$ and radii between 10 and 100~pc \citep{2017A&A...601A.146C}.  
{We recall that for the existing GMC catalog  and in the rest of this paper, the CO-to-H$_2$ conversion factor used is constant and equal to  
4$\times 10^{20}$~K$^{-1}$~km$^{-1}$~s~cm$^{-2}$ unless stated otherwise}. The estimated mass of 490 clouds is  above the survey completeness limit of 6.3$\times 10^4$~M$_\odot$.  
\citet{2018A&A...612A..51B} have shown that the GMC mass spectrum  becomes steeper  moving towards large galactocentric radii, and  the maximum mass of GMCs is also a factor 2 smaller
than in the inner disk.  A similar trend has been found for the 24~$\mu$m flux of the MIR source in the M33 area  by \citet{2011A&A...534A..96S}:   the cumulative distribution  
steepens in the intermediate disk,  where  the maximum source luminosity is lower than in the inner disk.  The ratio of molecular cloud mass to MIR flux density at 24~$\mu$m
does not show a clear radial trend: it has  a mean value of about 4$\times 10^4$~M$_\odot$~mJy$^{-1}$ and a large dispersion (of about 0.7 dex) at all radii. 

\begin{figure}
\centering
\includegraphics[width=9cm]{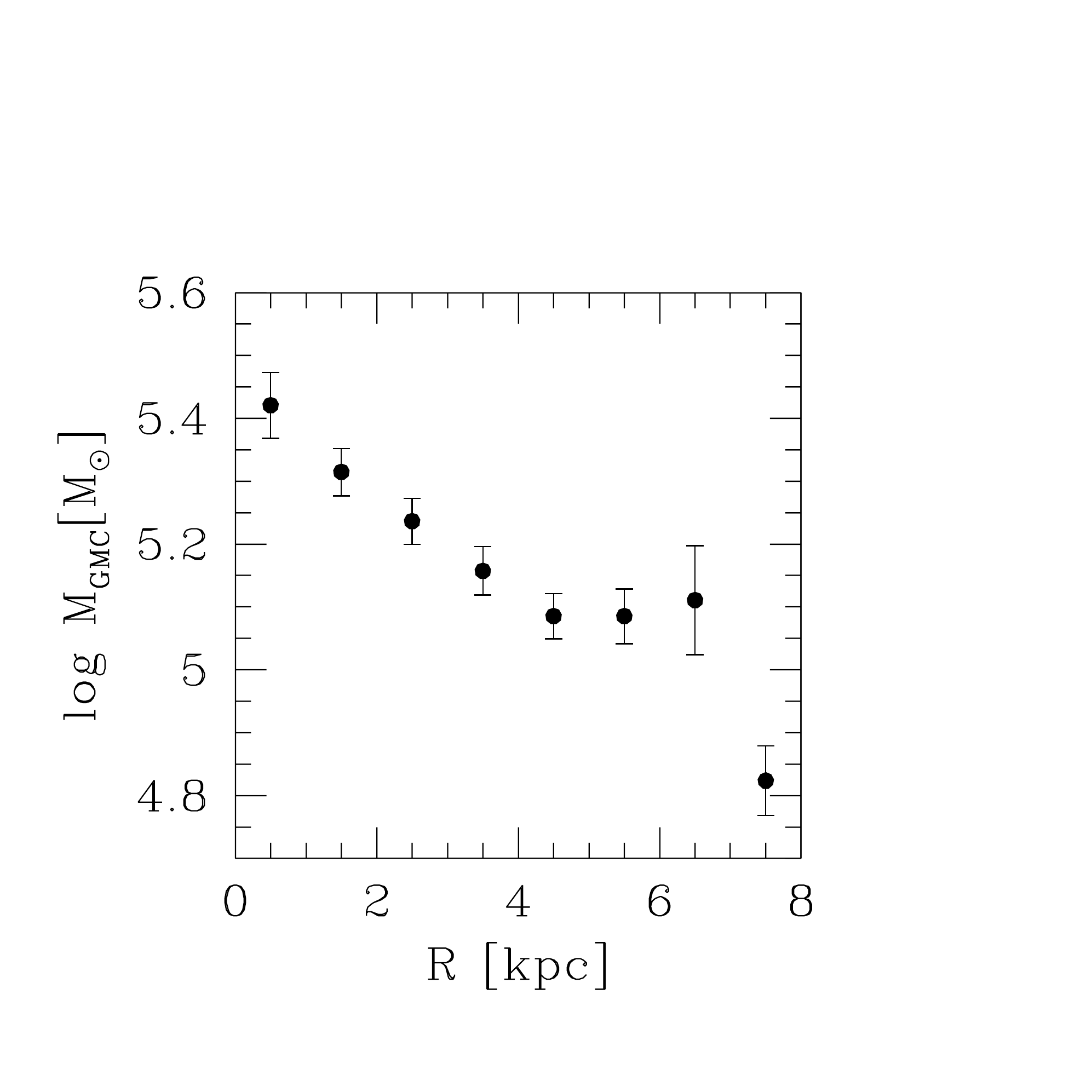}
 \caption{Mean value of GMC mass in dex in radial  bins of galactocentric distance  from data in the GMC catalog of \citet{2017A&A...601A.146C}.}
\label{massb}
\end{figure}

The number  of GMCs per unit area decreases continuously with radius across the disk. However, the average mass of GMCs decreases radially in the inner disk  (which hosts 410 GMCs), but then it
flattens in the intermediate disk (hosting 152 GMCs) and drops in the outer disk (where only 4 GMCs are found since the all-disk survey does not cover much of the outer disk). 
Figure~\ref{massb} shows  this radial trend. The  analysis of \citet{2017A&A...600A..27G}, dedicated to investigating whether and how CO traces  
molecular cloud masses in the SF disk, confirms  the rather constant fraction of dark gas  across the SF disk that is, that the CO-to-H$_2$ conversion factor does not increase 
at large galactocentric radii in the SF disk.  
This ensures that the observed trend for the mean molecular mass in the inner and intermediate disk, inferred for a constant CO-to-H$_2$ conversion facto, is indeed real.  The increase in mean GMC 
mass in the inner disk  can be caused by the rotation of the spiral arm pattern, which collects clouds into more massive complexes inside 4~kpc. This mechanism can break beyond a certain
radius, the corotation radius. Farther out,
beyond 6.5~kpc, the disk is stable according to the Toomre stability criterion \citep{2003MNRAS.342..199C}. Alternatively, the shear may break the perturbations apart at
large radii and leave only GMCs below a certain mass. We examine these possibilities in Section~5 after discussing the results of additional  CO observations of selected targets in the outer disk
 in the next sections. 
 
\begin{figure}
\centering
\includegraphics[width=9cm]{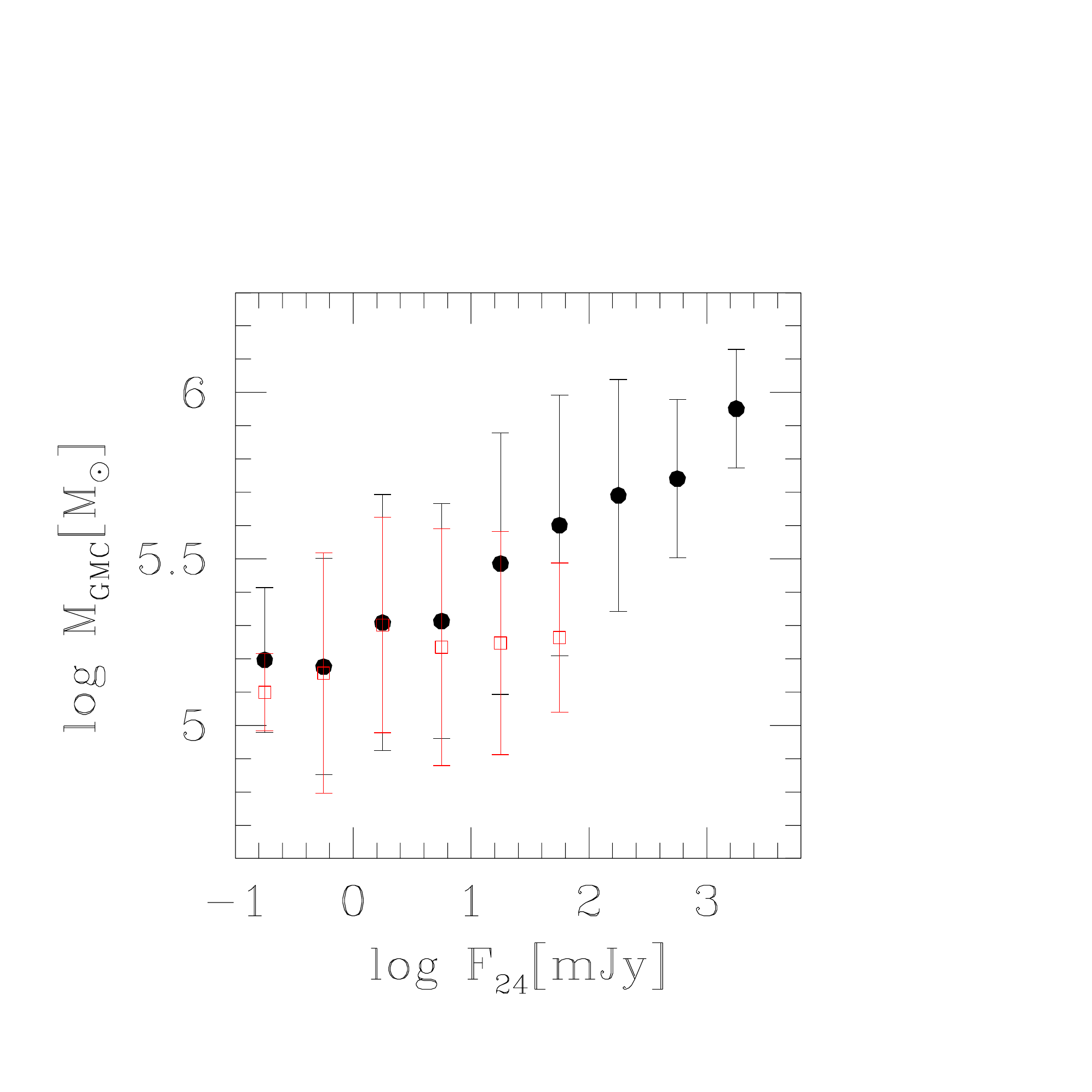}
 \caption{Mean value of  GMC mass hosting MIR sources  in the SF disk of M33 (black filled circles) in dex and in bins of F$_{24}$, the 24~$\mu$m source flux.
 Red open squares refer to GMCs and MIR sources in the intermediate disk.}
\label{gmc24}
\end{figure}

The mean mass of  molecular clouds hosting  MIR sources is weakly related to  the intensity of the 24~$\mu$m flux, F$_{24}$, for  bright sources. This is shown in Figure~\ref{gmc24}, where we compute
the mean GMC mass in bins of 24~$\mu$m flux intensity. For F$_{24}>5$~mJy, the mean GMC mass  increases as the 24~$\mu$m flux of the hosted source increases. However, this trend has a large scatter, and furthermore, it disappears for faint MIR sources or if we consider only 
MIR sources in the intermediate disk, which is  devoid of bright MIR sources (open red squares in Figure~\ref{gmc24}). This can be due to the limited sensitivity and completeness of the CO all-disk survey, and it needs further 
investigation by deeper CO searches around faint MIR sources. However, when we extrapolate the trend observed for F$_{24}>5$~mJy, the expected mean mass of a GMC hosting an MIR source with F$_{24}\simeq$ 0.1-1~mJy 
is about 10$^5$~M$_\odot$, only one order of magnitude below the mean mass of GMCs that are associated with the brightest  MIR sources  ( F$_{24}\sim$ 10$^3$~mJy). It is then conceivable to detect molecular gas associated
with faint MIR sources through pointed observations, even though the CO-to-H$_2$ ratio  might decrease in the outer disk.
Molecular line emission has been detected  in the SF disk of M33  by \citet{2011A&A...528A.116C} at the location of compact MIR sources with 3$\le F_{24}\le$21~mJy  at the level of 0.3~K~km~s$^{-1}$. 
Estimated molecular cloud masses range between 10$^4$ and 10$^5$~M$_\odot$.
The detected lines suggest  that  low-mass  GMCs  might be ubiquitous around MIR sources  in the SF disk of M33, despite the large spread in CO-to-F$_{24}$ flux ratio.
This needs to be examined  for fainter MIR sources and   in the outer disk.  

In the rest of this section, we select a faint MIR sample in the outer disk for deep pointed CO observations. These observations  are presented  in the next section, where
we additionally  describe the results of  a CO J=2-1 line map around a bright H$\alpha$ source in the outer disk.

\subsection{Multiwavelength data}

In analyzing the data of M33, we assumed a distance of 840~kpc  \citep{1991ApJ...372..455F,2013ApJ...773...69G},
which implies a linear scale of 4.1~pc per arcsecond. 
 
Dust emission at MIR wavelengths has been investigated through
the  InfraRed Array Camera (IRAC) and 
the Multiband Imaging Photometer (MIPS) on board {\it Spitzer}. 
The complete set of IRAC (3.6, 4.5, 5.8, and 8.0~\mi) and 
MIPS (24, and 70~\mi) images of \mm\ is described by \citet{2007A&A...476.1161V}.
To investigate the ultraviolet (UV) continuum emission of \mm, we used 
{\it Galaxy Evolution Explorer (GALEX)} data,
in particular those distributed by \citet{2007ApJS..173..185G}. 
To trace ionized gas, 
we adopted the narrow-line \Ha\ image of \mm\ obtained by \citet{1998PhDT........16G}
and described in detail in \citet{2000ApJ...541..597H}. 

The H$\alpha$ and FUV surface brightnesses in M33 decrease radially with a scale length of about 2~kpc out to about 6.5~kpc
\citep{2009A&A...493..453V}. Beyond this radius, they experience a sharper radial decline  (see also \citet{1989ApJ...344..685K}). The distribution of  the 
neutral atomic gas traced by 21cm line emission in M33 has been mapped by combining  
Very Large Array (VLA) and the Green Bank Telescope (GBT), and results are described by \citet{2014A&A...572A..23C}.
The atomic gas distribution starts to warp beyond the SF disk,  and the surface mass density  drops at R$_{25}$ (which is about 8.6~kpc), where the warp
becomes severe. We used the  tilted ring model fitted to the 21cm velocity field \citep{2014A&A...572A..23C}  to
estimate  galactocentric distances for sources in the warped outer disk.  

\subsection{Selection of the MIR sample in the outer disk}

From the catalog of 912 MIR sources in the M33 area \citep{2011A&A...534A..96S}, we  selected 99 sources that might be associated with SF sites  at galactocentric distances
 7.5$<$R$<$10.8~kpc and that have not been listed as M33 variable stars 
or have a Milky Way star as optical counterpart. The inner radial cutoff
was chosen to avoid overlaps with the  M33 all-disk CO J=2-1 survey covering the  SF disk \citep{2014A&A...567A.118D}, while the outer radius was dictated by the full coverage of 
Spitzer-IRAC maps of M33  and of other available  maps (GALEX, H$\alpha$). The analysis of CO observations of a previous sample of faint, 
compact MIR sources in the SF disk of M33  has revealed the importance of IRAC and MIR color-color diagram as a diagnostic tool to distinguish MIR sources 
that are truly SF sites  from evolved stellar objects, such as AGB stars and background 
galaxies \citep{2011A&A...528A.116C}. In the SF disk,   CO  J=1-0 and 2-1 lines have been detected around all MIR
sources    with  a characteristic spectral energy distribution between 3.6 and 24~$\mu$m  (see their Fig. 5).   The main-beam integrated  temperature for the CO J=2-1 lines 
for the selected MIR sources in the SF disk is between 0.7 and 2~K~km~s$^{-1}$ 
with peaks   between 0.1 and 0.5~K   \citep{2011A&A...528A.116C}  that are uncorrelated with the intensity of 24~$\mu$m flux  (in the range 1-21~mJy). 

\begin{figure}
\centering
\includegraphics[width=9cm]{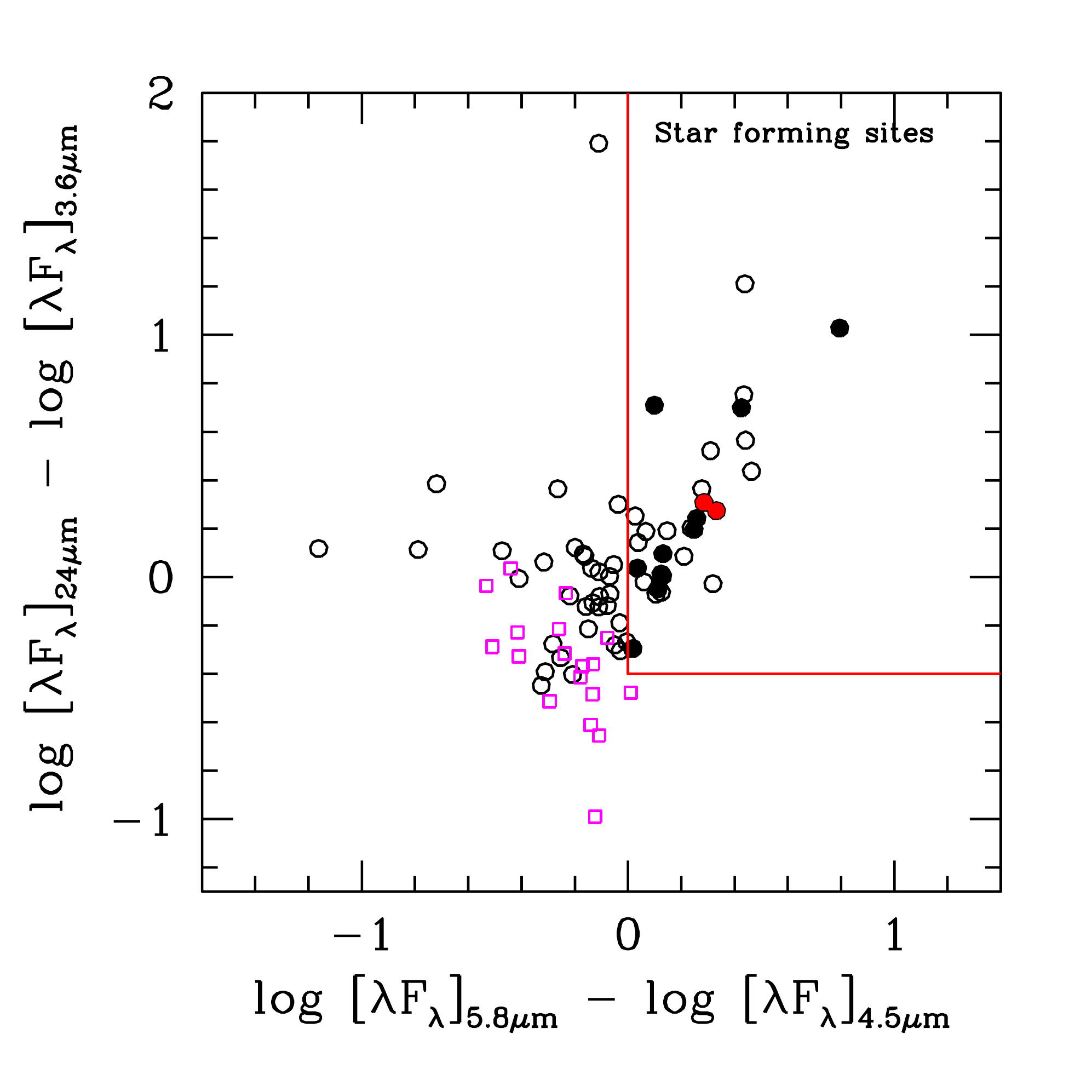}
 \caption{Colors or differences in energy densities at two wavelengths for MIR sources in the  catalog of \citet{2011A&A...534A..96S} that are at galactocentric distances 
between 7.5 and 10.8~kpc. The red box selects 30 MIR sources that are likely to be YSCCs. The filled symbols indicate the YSCCs selected for pilot observations of the CO J=1-0 and J=2-1 
lines presented in this paper. Filled red symbols   show the two YSCCs where the CO lines have been detected. The open magenta squares indicate the location of MIR sources that optically appear 
as background disk galaxies.}
\label{colors}
\end{figure}

To estimate the energy density in  IRAC bands,  we performed aperture photometry for the 99 selected sources with a fixed aperture size of 8~arcsec. We refer to  \citet{2011A&A...534A..96S}
 for  details on the photometry at 24~$\mu$m, in  H$\alpha$, FUV, and NUV bands and  for the relative uncertainties.
In the red box in the upper right corner of Figure~\ref{colors} we show the  30 MIR sources that have  [5.8]-[4.5]$> 0$ and [24]-[3.6]$>-0.4$ flux ratios and are our best YSCCs
beyond the edge of the SF disk of M33.  
Using the optical Sloan Digital Sky Survey images, we tagged 18 of the 99 sources that have an optical counterpart that is shaped as a disk galaxy. 
The 18 tagged sources, which are likely background disk galaxies, are plotted using open magenta squares.
Figure~\ref{colors} shows that none of them  lie in the selected region (red box). However, we cannot completely exclude contamination from
background galaxies, which are unresolved by the SDSS survey or are not disk like, and we discuss this possibility  further in Section~3.  For the pilot observations presented here, 
we  selected  12 of the 30 YSCCs  as a representative sample, and we call them  the MIR-selected sample. 

\begin{table*}
\caption{Properties of the MIR-selected sample.}
\label{phottab}
\begin{minipage}{\textwidth}
\begin{center}
\begin{tabular}{lcccccccccc}
\hline \hline
     ID    & RA  &  DEC   &  $r$    &  F$_{24}$ &  F$_{8}$ & log  L$_{H\alpha}$ & log L$_{FUV}$      & log L$_{NUV}$     & R \\
     ....   & deg  & deg   &  arcsec &  mJy        & mJy         & erg~s$^{-1}$        & erg~s$^{-1}$     &  erg~s$^{-1}$      & kpc \\
\hline \hline

 s537 & 23.6099 & 29.9906 & 1.8         & 1.29$\pm$0.02 & 0.14$\pm$0.02 &     .... &    .... &    .... & 10.4    \\ 
 s542 & 23.6651 & 30.0625 & 1.7     & 1.31$\pm$0.02     & 0.48$\pm$0.03  &     .... &    .... &    .... & 10.0       \\ 
 s555 & 23.2454 & 30.1715 & 1.8         & 1.07$\pm$0.02 & 0.33$\pm$0.03 &     ....  &    .... &    .... & 7.8       \\
 s631 & 23.1010 & 30.4421 & 1.3         & 0.39$\pm$0.02 & 0.35$\pm$0.03 &        ....  & 36.6 & 36.3    & 7.9       \\
 s672 & 23.8356 & 30.5626 & 1.6         & 0.83$\pm$0.02 & 0.73$\pm$0.04 &         .... &    .... &    .... & 8.9       \\
 s771 & 23.0613 & 30.7235 & 1.7         & 0.87$\pm$0.02 & 0.49$\pm$0.03 &         .... &    .... &    .... & 9.2       \\
 s787 & 23.8474 & 30.7463 & 2.0         & 2.62$\pm$0.02 &  ....  $\pm$....          &     ....  &   .... &    ....  & 8.5       \\
 s799 & 23.1572 & 30.7673 & 1.4         & 0.54$\pm$0.01 & 0.33$\pm$0.03  &    34.7 & 37.3 & 37.0 & 7.8        \\
 s854 & 23.2726 & 30.9742 & 1.4         & 0.45$\pm$0.01 & 0.26$\pm$0.02  &   ......   & 36.8 & 36.9  & 7.7        \\
 s892 & 23.7057 & 31.1582 & 1.4         & 0.45$\pm$0.01 & 0.37$\pm$0.03  &    35.7 & 37.8 & 37.5  & 8.1        \\
 s905 & 23.6735 & 31.2885 & 1.9     & 2.77$\pm$0.02 & 1.80$\pm$0.06  &        .... &    .... &    .... & 10.1        \\
 s907 & 23.4811 & 31.3279 & 2.0         & 1.94$\pm$0.02 & 0.43$\pm$0.03  &        .... &    .... &    .... & 9.8        \\

\hline \hline
\end{tabular}
\end{center}
\end{minipage}
\tablefoot{The identification number of each source in the MIR-selected sample is listed in Col. 1, and
the relative celestial coordinates are presented in Cols. 2 and 3. The estimated  source radius is given in Col. 4 and the flux at 24~$\mu$m  in Col. 5. The 8~$\mu$m flux is listed in Col. 6,  and the source luminosity in the H$\alpha$ line 
and in the near and far continuum UV are given in Cols. 7,8, and 9, respectively. In the last column we show the source galactocentric distance.}
\end{table*}

In Table~1 we list the identification number of each source in the MIR-selected sample according to the catalog of \citet{2011A&A...534A..96S},
its celestial coordinates, the estimated  source radius, and the flux at 24~$\mu$m. We also show the 8~$\mu$m flux and the source luminosity in the H$\alpha$ line and in the near and far continuum UV
if sources belong to M33. The galactocentric distance R,  estimated using the tilted ring model, is given in the last column.
Only 4 of the 12 sources have  detectable FUV emission with luminosities higher than 10$^{36}$~erg~s$^{-1}$
, and only two have a weak H$\alpha$ counterpart. While the lack of H$\alpha$ is expected for low-mass YSCC because of the  IMF incompleteness \citep{2009A&A...495..479C}, 
the paucity of FUV emission suggests that some source  
 might be associated with embedded YSCCs,  distant background galaxies, or QSOs. Most of the 18 sources that are tagged as disk galaxies have detectable FUV emission.

\begin{figure*}
\centerline{
\hspace*{3.5cm}\includegraphics[width=14cm]{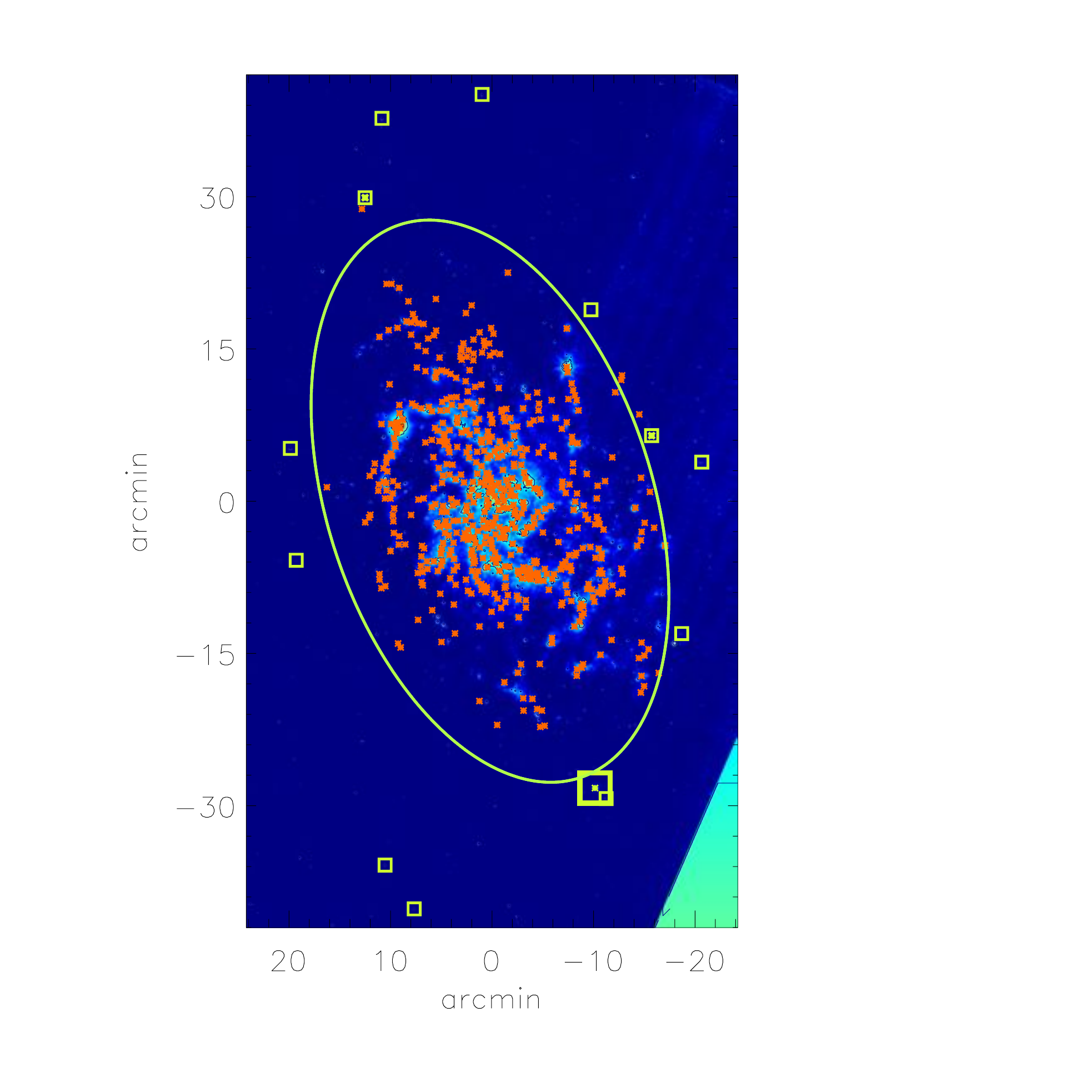}
\hspace*{-5cm}\includegraphics[width=14cm]{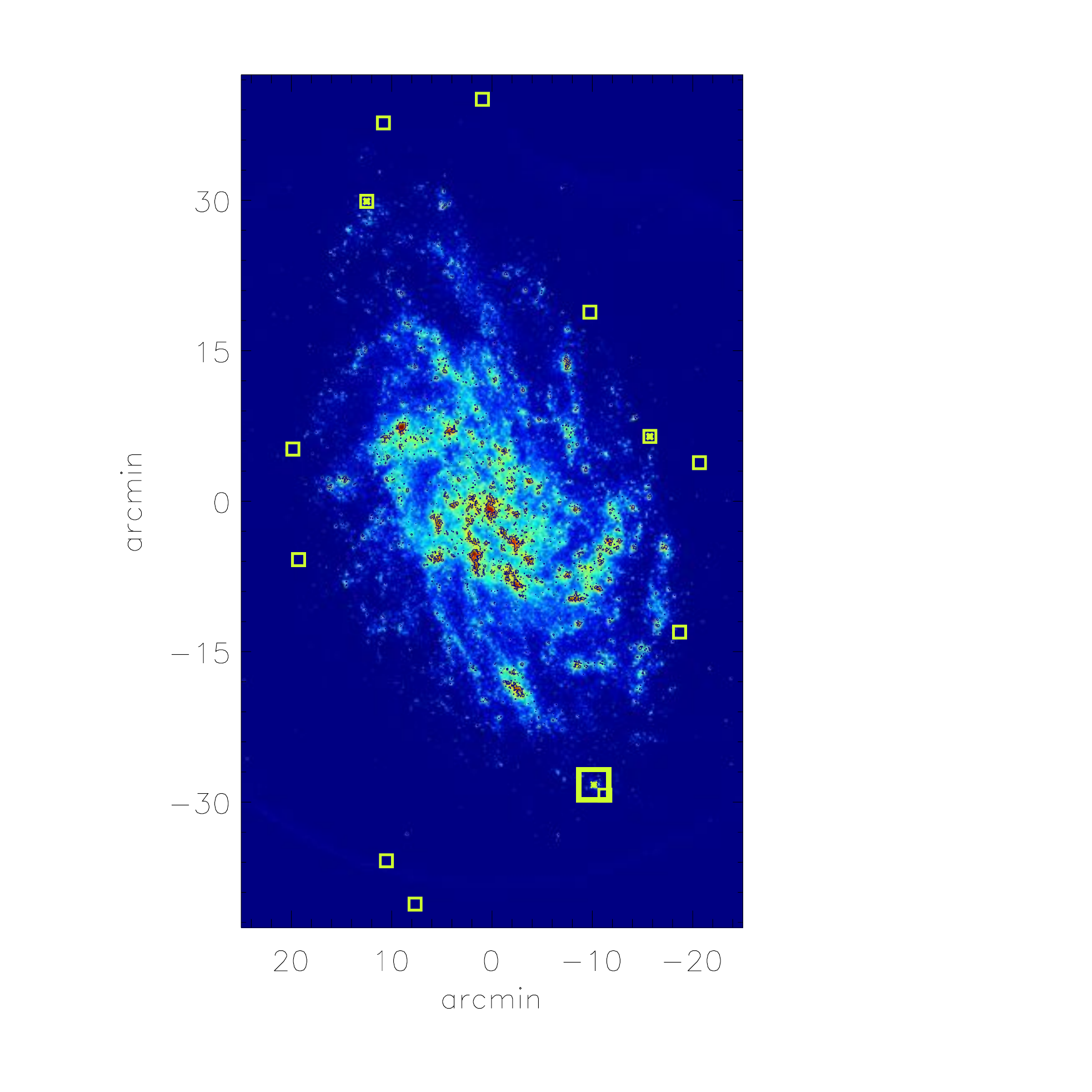}}
\caption{GMCs detected through the all-disk IRAM survey are plotted with orange asterisks over the 24~$\mu$m image of M33 in the {\it left panel}. The  
ellipse encloses the  SF disk, with a radius of 7~kpc,  mapped in the CO J=2-1 line by the IRAM all-disk survey.
Small open squares indicate the selected MIR sources  for the deeper CO pointed observations described in this paper, and the large open square indicates  the mapped southern field. 
Asterisks inside open squares highlight the location of the new detected molecular clouds presented in this paper. They are
plotted over the 24~$\mu$m image of M33 in the {\it left panel} and over the FUV-GALEX image of M33 in the {\it right panel}.  
}
\label{mirfuv}
\end{figure*}

In Figure~\ref{mirfuv} we show the location of the MIR sample over the 24~$\mu$m and FUV maps of M33. The squares indicate the  MIR sources that were selected for pointed
CO observations, while the asterisk in the squares highlights sources with detected CO emission  as described in the next section. 
In the left panel of  Figure~\ref{mirfuv}  we use  orange asterisks to indicate the location of the GMCs in the SF disk. The  ellipse encloses the SF disk area and
 has been drawn for  PA=23~$^\circ$ and $i$=54~$^\circ$  \citep{1991rc3..book.....D}. Selected sources may have
galactocentric distances that differ from what the ellipse position  suggests because of the change in orientation of the disk beyond 7~kpc.  

\begin{figure*}
\centering
\includegraphics[width=16cm]{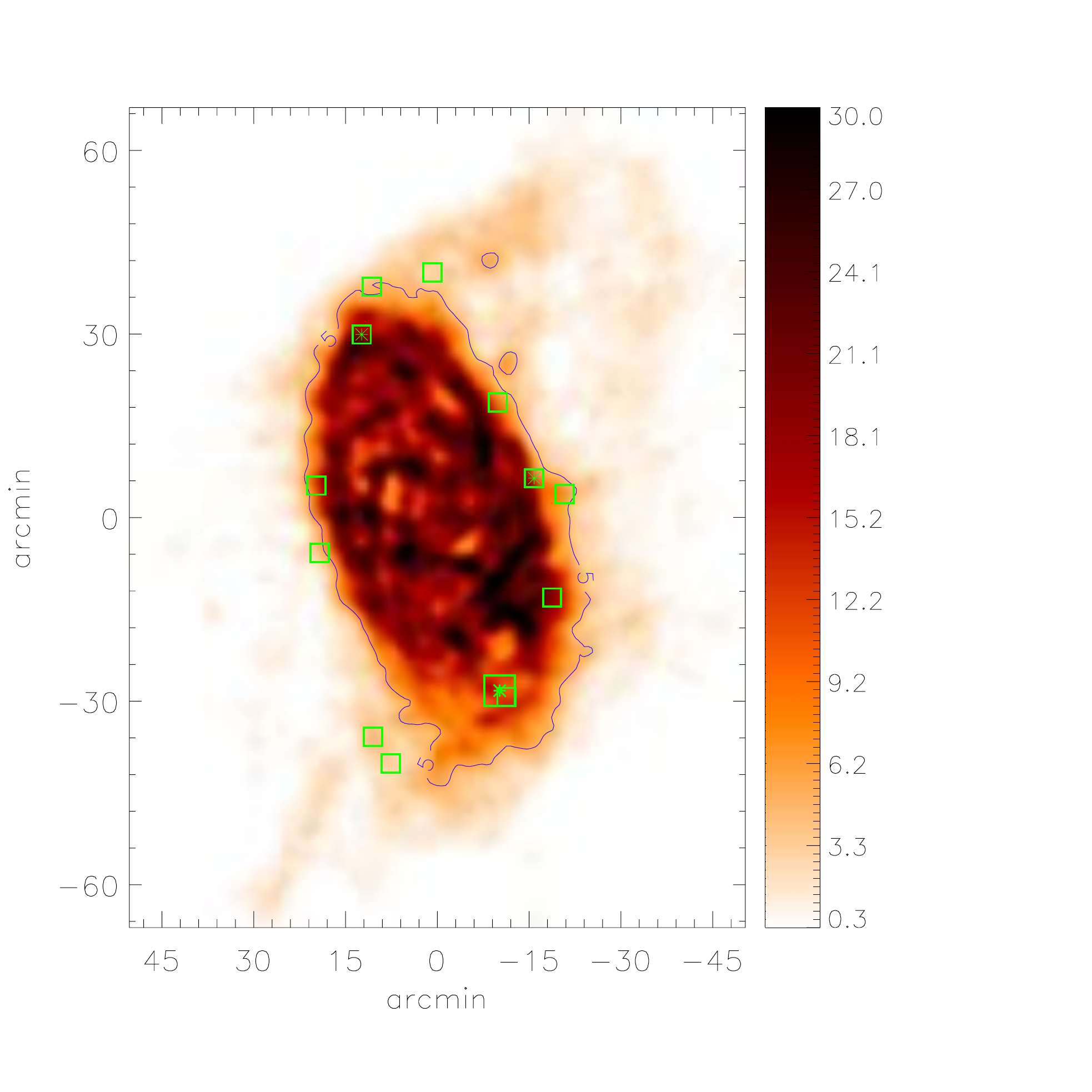}
 \caption{Location of sources in the MIR-selected sample (open squares) plotted over the 21cm map of M33. Asterisks indicate the
location of  MIR sources where CO has been detected. The color bar and map unit are Jy~beam$^{-1}$~km~s$^{-1}$; a 130~arcsec beam has been used for this smoothed 21cm map.
The dark contour level at 5~Jy~beam$^{-1}$~km~s$^{-1}$  corresponds to a face-on HI column density of about 2$\times$ 10$^{20}$~cm$^{-2}$.}

\label{sou2}
\end{figure*}

In Figure~\ref{sou2} we plot  the location of  selected  MIR sources over a smoothed version of the 21cm line map (equivalent beam FWHM=130~arcsec$^2$)
in order to recover emission from the outer disk. The atomic gas map shows that even though all sources are located beyond the SF edge of M33,
the majority of them still overlap the bright atomic disk,
before the 21cm surface brightness drops and the faint warped outer disk takes place.  
The dark contour level is at 5~Jy~beam$^{-1}$~km~s$^{-1}$ , which correspond to a
face-on value column density of about 2$\times$ 10$^{20}$~cm$^{-2}$.  Most of the sources  lie in regions of local HI overdensities, which can shield
molecules from the weak interstellar dissociating radiation field. Metallicities  (expressed as 12+log(O/H)) are expected to be 
 roughly a factor 3-4 below solar at galactocentric distances of the sources  and imply a  CO-to-H$_2$ conversion factor larger than the Galactic in the solar neighborhood 
 \citep{2016A&A...588A..23A}, that is, between 1.3 to 2 times the
 conversion factor used for the  SF disk  (which reads X$_{CO}$= 4$\times 10^{20}$~K$^{-1}$~km$^{-1}$~s~cm$^{-2}$  , \citet{2017A&A...600A..27G}).

\section{CO observations}
 
In this section we describe the   observations  of CO J=1-0 and CO J=2-1 lines  at the position of the selected MIR sources in the outer disk. 
We also present a 5~arcmin$^2$ map of the CO J=2--1 emission  around a bright HII region in  the southern side of the outer disk. 
We refer to this area as the southern field.   
 
\subsection{Pointed observations of the  MIR-selected sample}
 
 We searched for CO emission using the IRAM-30~m telescope  
at the location of  all sources in Table 1. The CO J=1-0 and J=2-1 lines were observed 
during March 2016 with  an FWHM beam of 21.4~arcsec at 115~GHz and of 10.7~arcsec at 230~GHz.  The  J=2-1 line is less likely to be affected by 
beam dilution than the J=1-0 line, but so far out in the disk, the excitation conditions may favor the lower transition.
 We observed all sources using the wobbler switching mode, making sure not to have CO emission in the reference beam.  The FTS  
backend with a spectral resolution of 195 kHz was used, corresponding to channel widths of  0.5~km~s$^{-1}$ at 115~GHz
and 0.25~km~s$^{-1}$ at 230~GHz. The VESPA backend system 
with  78.1 kHz resolution (0.2~km~s$^{-1}$) was also used at 115~GHz,  but the noise level was 
higher than for the FTS backend when data were smoothed at the same spectral resolution, therefore we only discuss the FTS data.  
All spectra were smoothed to a spectral resolution of 0.5~km~s$^{-1}$ , and data from both polarizations were  averaged. 
The beam and forward efficiencies  used to convert antenna temperatures into main-beam temperatures are
 0.71 and 0.95 at 115~GHz, respectively, and 0.65 and 0.92 at 230~GHz, respectively.  
 
Table 2 summarizes the CO J=1-0 and J=2-1 data. The identification number of the MIR source in each pointed observation  is in given in Col. 1. In Cols. 2 and 3 we show  the integrated 
line brightness per beam I of the J=1-0 and J=2-1 lines, respectively, in main-beam temperature units. 
In Cols. 4 and 5 we display the mean velocity V of  the two lines, and in Cols. 6 and 7, we list the relative line widths W. The peak intensity value P is given in Cols. 8 and 9, 
and  the rms noise level for the CO J=1-0 and J=2-1 spectra, smoothed at 2~km~s$^{-1}$ resolution, is presented in  Cols. 10 and 11, respectively. 
The I, V, and W values are those of the  best Gaussian fit to the line after smoothing the data to 
a resolution of 0.5~km~s$^{-1}$.  The integrated line brightness 
obtained by summing the flux in each channel inside the signal spectral window is given in parentheses, and it can be seen that the difference with the integral of the Gaussian fitted line is 
on the same order as  the fit uncertainty.
Using the rms noise,  we have estimated 4~rms upper limits for the integrated line emission at locations of sources with no detectable signal. The values quoted in Table 2
 are for a   8~km~$s^{-1}$ spectral window.  
 By requiring a 4~rms detection and for a typical rms value of 5 mK, we  have that I$_{1-0}< 80$~ mK~km~s$^{-1}$. Similarly,
using a typical rms value of 4~mK for the CO J=2-1 spectrum, we have that I$_{2-1}< 64$~ mK~km~s$^{-1}$ for undetected emission. 
The individual upper limits of the integrated line brightness for each observed position are given in Table 2. 

\begin{figure}
\centering
\includegraphics[width=9cm]{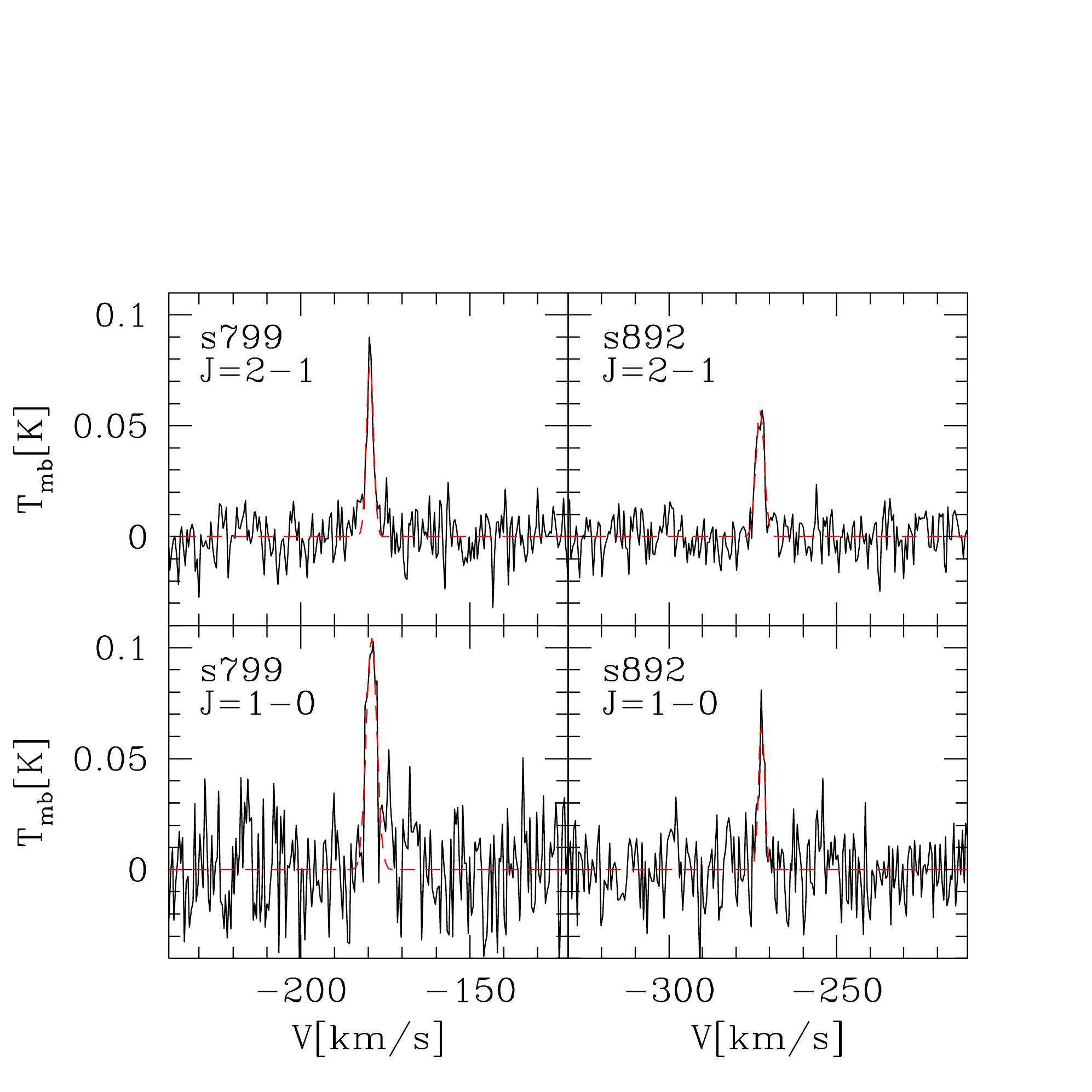}
 \caption{CO line emission detected at the location of sources s799 and s892 and the best-fit Gaussian function. Velocities are heliocentric, and the line emission is in main-beam
 temperature units.  }
\label{spectra}
\end{figure}

In Figure~\ref{spectra} we show the detected lines. Only 2 out of the 12 MIR sources were detected, and the peak velocities of the J=1-0 and J=2-1 correspond to within 0.5~km~s$^{-1}$ , therefore we can assume that the two lines come from the same cloud. The lines are narrow,
and  the  J=2-1 to the J=1-0 integrated line brightness ratio varies: it is 0.5 for s799 and 1.2 for s892. 
The small size of the MIR source and the high J=2-1/1-0 ratio with respect to the standard 0.8 value \citep{2014A&A...567A.118D}  suggests that beam dilution affects source s892.
On the other hand, the weakness of the radiation field in these outer
regions can easily cause the CO J=2-1 line to be dimmer and lower the J=2-1 to J=1-0 ratio,  as is observed for source s799 (see next section for more details on models for the detected gas).

We also pointed the telescope at five positions close to source s892 (s892-o1, o2, o3, o4, and o5) whose offsets in arcsec for RA and DEC are (11.5, -1.0), (-10.5,-11.4), (18, 11.4), (-36, 22.5), and (63,36.2),
and one position close to source s555 (s555-o1), whose offset is (7,-7).  No lines were detected in any of the offset locations.
The upper limits are on the same order as  the fluxes detected on source at (0,0) offset, and we conclude that  the  cloud  associated with source s892 does not have  brighter peaks 
in nearby locations.  This is supported by the null detection after stacking the offset spectra, as discussed in the next section.
The position of source s555  is in the lower right corner of the southern field, as discussed in the next subsection, and
its upper limit confirms that the emission in the southern field is not  very extended.

\begin{table*}
\caption{CO line pointed observations of the MIR-selected sample and Gaussian fits to detected lines.}
\label{lines}
\begin{minipage}{\textwidth}
\begin{center}
\begin{tabular}{lccccccccccc}
\hline \hline
ID     & I$_{1-0}$  & I$_{2-1}$ &  V$_{1-0}$  & V$_{2-1}$ &  W$_{1-0}$  & W$_{2-1}$ & P$_{1-0}$  & P$_{2-1}$ 
      & rms$_{1-0}$ & rms$_{2-1}$ \\
....  & mK~km~s$^{-1}$ & mK~km~s$^{-1}$ &  km~s$^{-1}$ &  km~s$^{-1}$ & km~s$^{-1}$ & km~s$^{-1}$ & mK & mK 
      & mK & mK \\
\hline \hline

 s537      &$<$ 77  &$<$ 49  & ....  & ....  & ....  & ..... & ....  & ....  &  4.8  &  3.1     \\ 
 s542      &$<$ 93 &$<$ 70  & ....  & ....  & ....  & ..... & ....  & ....  &  5.8  &  4.3  \\ 
 s555      &$<$ 75  &$<$ 75  & ....  & ....  & ....  & ..... & ....  & ....  &  4.7  &  4.7      \\
 s555-o1   &$<$ 140 &$<$ 117 & ....  & ....  & ....  & ..... & ....  & ....  &  8.7  &  7.3      \\
 s631      & $<$ 75 &$<$ 43  & ....  & ....  & ....  & ..... & ....  & ....  &  4.7  &  2.7      \\
 s672      & $<$ 97 &$<$ 80 & ....  & ....  & ....  & ..... & ....  & ....  &  6.1  &  5.0       \\
 s771      & $<$ 83 &$<$ 83  & ....  & ....  & ....  & ..... & ....  & ....  &  5.2  &  5.2       \\
 s787      & $<$ 77 &$<$ 57  & ....  & ....  & ....  & ..... & ....  & ....  &  4.8  &  3.6    \\
 s799     & 408$\pm45$ (467) &202$\pm 19$ (227) & -179.1  & -179.5  & 3.7  & 2.4 & 104 & 79 &  8.8 &  4.8  \\
 s854      & $<$ 87 &$<$ 89 & ....  & ....  & ....  & ..... & ....  & ....   &  5.4 &  5.6      \\
 s892     & 150$\pm23$ (149) & 179$\pm 11$ (171)  & -272.4  & -272.8  & 2.2 & 3.0 & 65 & 57 &  6.1 &  4.5   \\
 s892-o1   &$<$ 171 &$<$ 128 & ....  & ....  & ....  & ..... & ....  & ....   & 10.7 &  8.0       \\
 s892-o2   &$<$ 134 &$<$ 140 & ....  & ....  & ....  & ..... & ....  & ....   &  8.4 &  8.7        \\
 s892-o3   &$<$ 158 &$<$ 132 & ....  & ....  & ....  & ..... & ....  & ....   &  9.9 &  8.3        \\
 s892-o4   &$<$ 144 &$<$ 146 & ....  & ....  & ....  & ..... & ....  & ....   &  9.0 &  9.1        \\
 s892-o5   &$<$ 205 &$<$ 131 & ....  & ....  & ....  & ..... & ....  & ....   & 12.8 &  8.2       \\
 s905      &$<$ 101 &$<$ 63  & ....  & ....  & ....  & ..... & ....  & ....   &  6.3 &  3.9       \\
 s907      &$<$ 85  &$<$ 61  & ....  & ....  & ....  & ..... & ....  & ....   &  5.3 &  3.8       \\
 
\hline \hline
\end{tabular}
\end{center}
\end{minipage}
\tablefoot{
The ID of the MIR source in each pointed observation  is given in Col. 1. In Cols. 2 and 3 we show I, the integrated line
brightness per beam of the J=1-0 and J=2-1 detected lines, respectively (in parentheses we add the values obtained by summing the flux in each channel) or their upper limits. 
In Cols. 4 and 5 we display the mean velocity V of  the detected lines, in Cols. 6 and 7 we list the relative line widths W, and in Cols. 8 and 9  the peak intensity values P. 
In the last two columns we show the rms noise level for the CO J=1-0 and J=2-1 spectra  smoothed at 2~km~s$^{-1}$ resolution.}
\end{table*}

\subsection{Map of the southern field}

A region centered at RA=01:33:06 Dec=30$^\circ$11'30"  (RA=23.2750$^\circ$ DEC=30.1917$^\circ$), close to the position of a large HII region in the extreme south of M33,    
was included in the CO J=2-1 observations of the all-disk survey \citep{2014A&A...567A.118D}. 
This region was observed separately 
because the  position of the  center of M33  in the sky exceeds the IRAM 83$^\circ$ elevation limit for about 40~minutes every day.  
Defining the southern region as a separate source enabled us to observe it during about 25 of the 
40 minutes that would have been lost each day (the remaining time was used to check pointing).  It was observed at 82.8$^\circ$ elevation on average and is slightly beyond the southern limit of the map 
presented in \citet{2014A&A...567A.118D}.  It was therefore not presented in \citet{2014A&A...567A.118D}.

The bright HII region in the southern field, also known as MA1 \citep{1942ApJ....95....5M}, lies at a galactocentric distance of 7.4~kpc, and it is one of the few HII regions that are located beyond the drop of the 
average H$\alpha$ surface brightness. 
The metal content of this outer HII region  has been investigated, and the most recent estimate for its oxygen abundance is 12+log(O/H)=8.28~dex 
\citep{2010A&A...512A..63M}. This HII region  is coincident with  the MIR source identified by 
\citet{2011A&A...534A..96S} with ID number 562, which has a  flux of 9.2~mJy and a size of  4.4~arcsec at 24~$\mu$m. It hosts a stellar cluster whose stellar mass and age have been
estimated to be about 6000~M$_\odot$ and 8~Myr  by \citet{2011A&A...534A..96S}. This HII region is much brighter than any other in the MIR  sample selected in the outer disk.
 
\begin{figure*} 
\centering
\includegraphics[width=15cm]{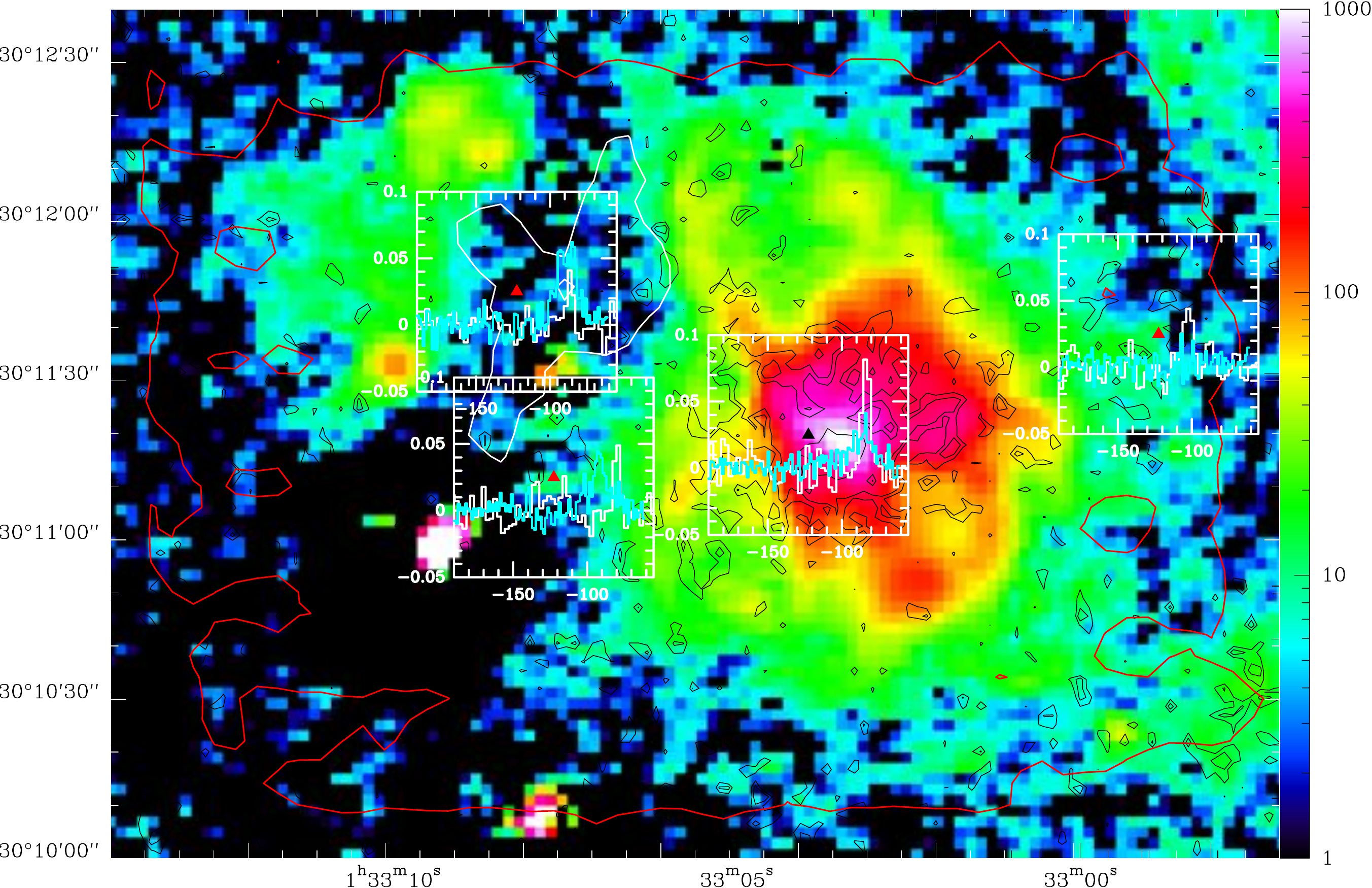}
\caption{Southern field H$\alpha$ emission in color, with  100$\mu$m PACS emission in black contours. Four CO spectra in the southern field where the signal is well above the noise are shown, 
with the precise positions indicated with red or black triangles.The 21cm spectrum at the location of the detected CO
line is shown in cyan. The thick red contour indicates the extent of the CO map.  
The white contour between the H$\alpha$ emitting zones shows the $N_{HI} = 1.8 \times 10^{21}$ cm$^{-2}$ column density level (in the rest of the region, the HI column densities are lower than the contour value).
 }
\label{south_r} 
\end{figure*}

The southern field of M33 was observed as described in \citet{2014A&A...567A.118D}, using on-the-fly mapping in CO(2--1) with the HERA multibeam receiver. 
Figure~\ref{south_r} shows the H$\alpha$ emission in color with the 100$\mu$m PACS emission in black contours and the positions where CO emission is well above the noise. 
The thick red contour indicates the extent of the CO map.  The four CO spectra  with a high enough signal-to-noise  ratio are overplotted, and the 
triangles indicate the corresponding position in the disk.  Spectra within 5~arcsec of the nominal position were summed to improve the signal-to-noise ratio. This broadens the beam 
only slightly since these spectra are all within the half-power size of a single beam. The nominal position of the spectra is given by the pixel with the highest signal-to-noise ratio. 
A white contour shows where  $N_{HI} = 1.8 \times 10^{21}$ cm$^{-2}$, that is, where the HI column density is high. 
In the southern field the HI column density peaks {\it \textup{in between}} the two SF regions and not where star formation is taking place, which is likely due  to  the transition from HI to H$_2$  and to the expansion of the
HII region shell, which compresses the surrounding neutral ISM  and thus enhances the local density. 

Each box in Figure~\ref{south_r} shows the CO(2--1) 
spectrum in white and the HI line temperature (divided by 1000) in light blue.  The CO spectra are presented on the  T$_{mb}$ scale and all spectra are in Kelvin.
The CO emission is strongest close to the HII region center.  All CO line widths are typical of single GMC, although they are broader than most spectra in the outer disk of M33, 
as for sources  s799 and s892  (but see also \citet{2012A&A...548A..52B}),  presumably due to the high level of star formation and to the evolution of the HII region.
Table~3 gives the integrated intensities and Gaussian fit results in main-beam temperature units.
The beam and forward efficiencies  used to convert antenna temperatures into main-beam temperatures at the time of these observations are 0.55 and 0.92 at 230~GHz. 
 The uncertainties are from the fitting program in CLASS (see http://www.iram.fr/IRAMFR/GILDAS/).
Uncertainties on the integrated intensities in Col. 2 can be calculated as  $\Delta I = 1.5 rms \times \sqrt{{\hbox{W}}_{2-1} \times 2.6}$ where rms is given in Col. 8 and W$_{2-1}$ in Col. 5.  
These uncertainties are similar to those in the GaussArea task in CLASS. We did not subtract the 2.6~km/s channel width, therefore the W$_{2-1}$  values in Col.~5 are upper limits.

\begin{table*}
\caption{Gaussian fits to the CO J=2-1  integrated line  brightness per beam for the detected emission  in the southern field. 
}
\label{lines2}
\begin{center}
\begin{tabular}{lccccccc}
\hline \hline
Name & Offset         &  I$_{2-1}$                 &   V$_{2-1}$             &   W$_{2-1}$    &  P$_{2-1}$  & $\Delta$V$_{2-1}$  & rms$_{2-1}$ \\
  ....    &  arcsec       & mK~km~s$^{-1}$   &  km~s$^{-1}$          &  km~s$^{-1}$ &  mK              & km~s$^{-1}$              & mK \\
\hline \hline
562a & (-562,-1704)   &  427$\pm84$(375)     & $-79.9\pm0.4$       & $4.3\pm1.1$   &  94    &  -83,-77          & 16.9 \\
562b & (-676,-1677)   &  595$\pm104$(492)   & $-102.4\pm0.6$     & $7.2\pm1.4$  &   77    &  -107,-98        & 16.1 \\
562c & (-610,-1696)   &  927$\pm100$(825)  &  $-83.6\pm0.3$      & $6.0\pm0.7$    &  144  &   -89,-80         & 17.2 \\
562d & (-555,-1669)   &  341$\pm67$(336)    & $-86.5\pm0.5$       &  $4.5\pm0.9$   &  70     &   -91,-82         &  13.4 \\
\hline \hline
\end{tabular}
\end{center}
\tablefoot{
The name of detected cloud clump is given in Col. 1. Column 2 gives the cloud clump offset with respect to the center of M33.
Column 3 gives the CO line intensity  in main-beam temperature units computed as the area of the best-fit Gaussian function,  
in parentheses we list the values obtained by summing the flux in each channel of the selected spectral window.
Columns 4 and 5 give the central velocity and the line width of the Gaussian fits, respectively.  The  signal peak and  spectral window are given in Cols. 6 and 7, respectively.
Column 8 gives the per channel noise level of the spectra (for a channel width of 2.6 km~s$^{-1}$).  
}
\end{table*}

\section{Radial decline of the  molecular cloud mass in M33}

In this section we  discuss the results of  detected and undetected CO line emission. We averaged   spectra of undetected sources to
further lower the rms and search for very faint CO lines in the outer disk. 
We  evaluated possible background contamination in the MIR sample, which might be severe  for faint sources
at large galactocentric distances, in order to understand the paucity of CO lines in the MIR sample. Finally, we
compare some characteristics of the HII region in the southern field, including
 the  CO line intensities, to those of SF sites  in the SF disk, which share a similar morphology  and  MIR emission. 
 This should help us  understand whether the outer disk 
 environment   or  the YSCC  evolution plays a major role in shaping the observed properties of MA1.

\subsection{Detected lines and the associated molecular clouds}

The molecular mass at the location of sources  s799 and s892 is uncertain because it depends on the extent and position of the clouds with respect to the beam. When clouds are smaller than the  FWHM
of the 115~GHz beam, as suggested by the lack of CO detection in the proximity of s892,
 the  measured CO J=2-1/1-0 integrated brightness temperature ratio is higher than  the intrinsic line ratio R$_{21}$,  defined as the J=2-1/J=1-0 line ratio for extended sources because beam dilution is stronger for the J=1-0 line.  Moreover, the CO-to-H$_2$ conversion factor can be different than the constant value  determined and used  for the SF disk.
 In what follows we estimate  cloud masses using different methods, assuming that  clouds are centered with respect to the beam and uniform in brightness.
 Using  Eqs. (5), (6), and (8)  in  \citet{2011A&A...528A.116C}, we can estimate  cloud sizes and masses from the observed J=2-1/1-0 line ratio, and 
 we refer to this method as the ratio method. In practice, we   convolved the cloud surface brightness with the beam for  the J=1-0 and J=2-1 line assuming a cloud size D and an intrinsic
 line ratio R$_{21}$. We then inferred D by  equating the expected line ratio to the observed one. The cloud mass was then estimated  using
 a CO-to-H$_2$ conversion factor,  X$_{CO}$,  which depends on metallicity. We  can also evaluate the cloud size and mass assuming virial equilibrium for a given  CO-to-H$_2$ 
 conversion factor. With this method, called   the virial {\it } method, the cloud size is determined by equating the virial mass to the luminous mass as in Eq.(18) of  \citet{2011A&A...528A.116C}. 
 When the J=1-0 and J=2-1 CO lines are both detected, R$_{21}$ is estimated.
 Finally, in the third method, the fill{\it } method, we assumed that the source is as extended as the FWHM beam at 230~GHz. We computed the mass assuming a CO-to-H$_2$ conversion factor 
  that depends on metallicity, and derived the cloud CO surface brightness by equating the observed CO J=2-1  line brightness to the expected one using beam convolution. 
 With this method,  the intrinsic line ratio R$_{21}$ was estimated only when the J=1-0 line was also  detected, otherwise   R$_{21}$ was an input value.
 
 We estimated the metallicity Z at the location of sources s799 and s892  using the galactocentric distance R and the oxygen abundance radial gradient of \citet{2010A&A...512A..63M}.  Following  \citet{2016A&A...588A..23A},
 we assumed a Z$^{-1.5}$ metallicity dependence of X$_{CO}$ and  write
 
 \begin{equation}
{\hbox {log}} {X_{CO}\over 2 \ 10^{20}} = 12.85-1.5(12+{\hbox{log}}{O\over H}) = 0.1+0.066\ R
 .\end{equation}
 
 For R=3~kpc, this formula  gives X$_{CO}$=  4$\ 10^{20}$~K$^{-1}$~km$^{-1}$~s~cm$^{-2}$, the  commonly used conversion factor for the SF disk of M33, and hence it  well approximates the
 average value of X$_{CO}$  derived for the inner and intermediate disk by
\citet{2017A&A...600A..27G}. These authors found  that for M33 the CO-to-H$_2$ conversion factor is independent of radius  out to about 6~kpc,   but beyond this radius, the lack of GMCs prevents any
definitive conclusion.
 
 We list in Table~4 the estimated molecular cloud parameters for the two detected sources in the MIR sample using the three methods.  Estimated cloud masses include   He and heavier elements.
 For the ratio method we used three different values of
 R$_{21}$ and quote the results when a value of the cloud size D satisfies the assumptions (i.e.,  that gives the observed J=1-0 and J=2-1 integrated line brightness for the assumed
 intrinsic ratio R$_{21}$ ).
 For source s892, all methods give similar cloud mass estimates, on the order of 2~$10^4$~M$_\odot$, while the  cloud extent is not well determined.
The low I$_{2-1}$/I$_{1-0}$ ratio observed at the location of source s799 might be due to the low-excitation conditions of the gas, because the SF source is weak in FUV and H$\alpha$ emission. 
At this location, the cloud mass can be as high as 10$^5$~M$_\odot$ and more extended than the beam FWHM at 115~GHz. Smaller sizes and lower masses are inferred using the virial or the fill method, which
predict low values of R$_{21}$.  
The expected stellar masses of  YSCCs associated with sources s799 and s892 are  200-300 M$_\odot$, as determined by the spectral energy distribution fits \citep{2011A&A...534A..96S}, and they
imply a    total stellar cluster-to-cloud mass ratio lower than 0.02.

The molecular cloud masses associated with sources s799 and s892, although very uncertain, confirm the paucity of giant molecular complexes beyond the SF edge and the decrease in mean value of  
molecular cloud  mass from the intermediate to the outer disk.  We underline that the results shown in Table~4 are relative to a CO-to-H$_2$ conversion factor that is  8.2 $10^{20}$ and 
8.6 $10^{20}$~K$^{-1}$~km$^{-1}$~s~cm$^{-2}$ for sources s799 and s892, respectively. These values are about a factor 2 higher than 
that inferred for the SF disk by \citet{2017A&A...600A..27G} and used by \citet{2017A&A...601A.146C}  and  by \citet{2018A&A...612A..51B}.  Although beyond the SF edge the conversion factor  may
increase  with respect to the SF disk as a result of more extreme physical and chemical conditions, it is unlikely that molecular clouds associated with sources s799 and s892 are much more massive than what we
quote in Table~3.  This implies that the average mass and size of perturbations that occasionally grow  in the outer disk
drop beyond the SF edge. The low  mass of the YSCCs  that are hosted by these clouds  provides additional support to this conclusion. 

Using the virial and fill method,  we also quote in Table~4 the cloud mass at the four positions listed in Table~3 in the southern field.  The metallicity was determined for  MA1 in the
southern field to be 12+log(O/H)=8.28 \citep{2010A&A...512A..63M}, and this implies a  CO-to-H$_2$ conversion factor X$_{CO}$=5.4 10$^{20}$~K$^{-1}$~km$^{-1}$~s~cm$^{-2}$, according to the above equation.
The strong radiation field of the HII region in the southern field suggests that very low values of R$_{21}$ are excluded. The total molecular hydrogen mass estimated for R$_{21}$=1.2 in
the southern field is very similar for both the virial and the fill method, being  5.4 10$^4$ and 6.1 10$^4$~M$_\odot$ , respectively. The total molecular hydrogen mass is in agreement with the mass of other 
GMCs measured in the outer disk and plotted in Figure~\ref{massb}. The virial method implies more compact clouds, but as discussed
later in this section, virial equilibrium is unlikely for gas in the proximity of an HII region that is not compact. Radiation from the stellar cluster breaks through the original cloud, which
is swept away by the expanding ionized shell. Given the linear scaling of  cloud mass with 1/R$_{21}$  for the fill methods, we estimate a  total molecular mass of about 10$^5$~M$_\odot$ for R$_{21}$=0.8,
which is still in agreement with a drop of the mean molecular cloud mass beyond the SF disk.

\begin{table}
\caption{Estimated properties of the detected molecular clouds in the outer disk for a  metallicity-dependent CO-to-H$_2$ conversion
factor.}
 \begin{tabular}{lccccc}
\hline \hline
     ID  & line & R$_{21}$ & D         & M                            &  Method \\
          &        &                 & arcsec & 10$^4$ M$_\odot$ &              \\
\hline \hline

799 & 2-1,1-0 & 0.4   &  34    &  9.9 &    Ratio   \\
799 & 1-0       & 0.15  &  9.2  &  5.1 &    Virial   \\
799 & 2-1       & 0.15  &  9.2  &  2.2 &    Virial   \\
799 & 2-1,1-0 & 0.26  &  11   &  5.2 &    Fill      \\
       &             &          &         &        &               \\
892 & 2-1,1-0 & 0.4   &  12    &  2.1 &    Ratio   \\
892 & 2-1,1-0 & 0.8   &  27    &  3. 0&    Ratio   \\
892 & 1-0       & 0.37  &  10   &  2.0 &    Virial   \\
892 & 2-1       & 0.32  &  5.2  &  1.9 &    Virial   \\
892 & 2-1,1-0 & 0.63  &  11   &  2.0 &    Fill      \\
       &             &          &         &        &               \\
562a  &   2-1   &  0.8   &  2.0    &  1.5      &   Virial     \\
562b  &   2-1   &  0.8   &  1.0    &   2.1     &    Virial      \\
562c  &   2-1   &  0.8   &  2.2    &   3.3     &   Virial     \\
562d  &   2-1   &  0.8   &  1.4    &   1.2     &    Virial     \\
       &             &          &         &        &               \\
562a  &   2-1  &   1.2  &   1.3     &   1.0     &  Virial     \\
562b  &   2-1  &   1.2  &   0.7     &    1.4    &  Virial      \\
562c  &   2-1  &   1.2  &   1.5     &    2.2    &  Virial      \\
562d  &   2-1  &   1.2  &   1.0     &    0.8    &  Virial      \\
       &             &          &         &        &               \\
562a   &  2-1   &   1.2      &  11      &  1.1    & Fill    \\
562b   &  2-1   &   1.2      &  11      &  1.6    & Fill   \\
562c   &  2-1   &   1.2      &  11      &  2.5    & Fill    \\
562d   &  2-1   &   1.2      &  11      &  0.9    & Fill   \\
\hline \hline
\end{tabular}
\tablefoot{The CO rotational line used to 
infer the cloud parameters is given in Col. 2, the intrinsic CO line ratio in Col. 3, and the
cloud diameter in Col. 4. The cloud mass is  shown in 
Col. 5, and   in the last column, we indicate the method used to derive the cloud parameters. }
\end{table}

\subsection{Stacking spectra}

The ratio of the integrated CO J=1-0 line brightness to the 24~$\mu$m source flux, I$_{1-0}$/F$_{24}$, is 0.76 and 0.33~K~km~s$^{-1}$~mJy$^{-1}$ for sources s799 and s892, respectively.  
These values are higher
than those in the sample of \citet{2011A&A...528A.116C}  for brighter MIR sources in the SF disk (0.05$<$ I$_{1-0}$/F$_{24}<$ 0.28~K~km~s$^{-1}$~mJy$^{-1}$).
 For  undetected sources of the MIR sample presented in this paper,  this ratio is always lower than 0.2~K~km~s$^{-1}$~mJy$^{-1}$ and hence 
 well below the values  for the two  detected sources  in the outer disk. To  further lower this limit, given the lower CO-to-24~$\mu$m flux ratios measured 
 by \citet{2011A&A...528A.116C} in the SF disk, we  stacked the CO spectra of MIR sources with no detected lines. 

\begin{figure}
\centering
\includegraphics[width=9cm]{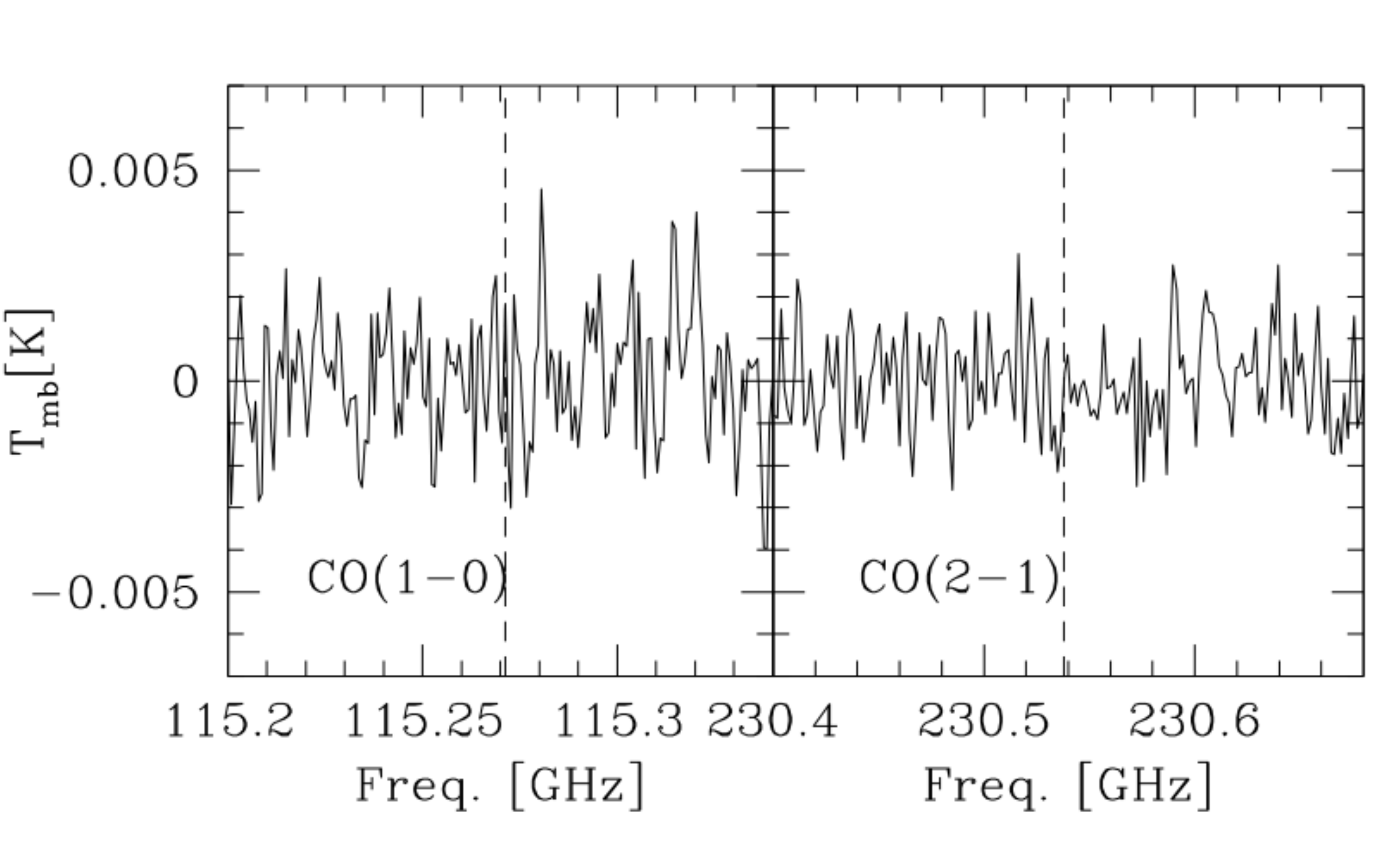}
 \caption{ Stacked spectra  obtained by averaging individual spectra of undetected sources after aligning them to  the rest frequency of the expected CO lines. 
 The vertical dotted line indicates the location in the spectra where a faint  CO J=1-0 or CO J=2-1 line  should be detected if some faint emission line were present in  individual spectra.}
\label{stack}
\end{figure}

We  applied the stacking procedure to all CO spectra with an
offset near source s892 and independently to the ten CO spectra with undetected line emission centered at the position of MIR sources. We performed this for the CO J=1-0 and J=2-1 line separately by 
computing the expected  frequency of the CO lines for each source. We aligned 
the spectra according to the expected frequencies and  averaged them. We  estimated the expected line frequencies from the
21cm line velocity because 21cm emission  is present at all selected positions in the disk. Despite the very low noise in the stacked spectra, no CO lines were detected. The  
stacked spectra relative to the ten undetected sources for the CO J=1-0 and J=2-1 lines are shown in Figure~\ref{stack}. The rms is as low as 1~mK for the J=2-1 line  
and  is 1.5~mK for the J=1-0 line. The dashed line in Figure~\ref{stack} indicates the location of the expected line. We have to underline   that the detected CO lines in the
MIR sample are very narrow and this increases the difficulty of the stacking technique  given the velocity resolution of 1.25~km~s$^{-1}$ in the HI database  and the
possible  shift of molecular gas velocities by a few km~s$^{-1}$ with respect to the velocities of the diffuse atomic gas.

The low rms of the stacked spectra implies that  for a 8~km~s$^{-1}$ spectral window, the average ratio of the integrated CO J=1-0 line brightness to the 24~$\mu$m source flux is
lower than 0.02~K~km~s$^{-1}$~mJy$^{-1}$. This upper limit is more than a factor 10 lower than the measured CO-to-24$\mu$m flux ratio for the two MIR sources  detected in the outer 
disk and lower than any measured CO-to-24$\mu$m flux ratio in the whole disk of M33. The lack of CO detection for most of the sources in the MIR sample
casts doubts on the association  of the same faint MIR sources  with the M33 disk, and we address this question in the next subsection.

\subsection{Background source count at 24~$\mu$m}

The MIR source catalog of \citet{2011A&A...534A..96S} contains  912  sources that lie in the sky area covered by the M33 gaseous disk. They have  24~$\mu$m fluxes above 0.2~mJy,  
and only 240 sources are brighter than 5~mJy. The area of the survey is approximately 0.75~deg$^2$, equivalent to 161~kpc$^2$ at the distance of M33. 
 At bright flux densities, F$_{24}\ge$ 5~mJy, the background source counts increase at approximately the Euclidean rate, 
and we expect only 23 sources in the survey area \citep{2004ApJS..154...70P}, that is,  a contamination by background sources  lower than 10$\%$. Between 0.2 and 5~mJy the MIR catalog 
might instead suffer significantly from contamination by background sources  because the number  density of these increases as the flux decreases. 
There is  evidence that  the number of background sources increases at super-Euclidean rate  below 5~mJy \citep{2004ApJS..154...70P}.
However, the MIR source catalog in the M33 area is 
far from being complete below 5~mJy because the sensitivity of the survey varies across the sky, to  crowding and diffuse emission in the inner regions of M33. Hence, both MIR
sources located in the M33 disk and background sources from more distant sources might not have been detected.

\begin{figure}
\centering
\includegraphics[width=9cm]{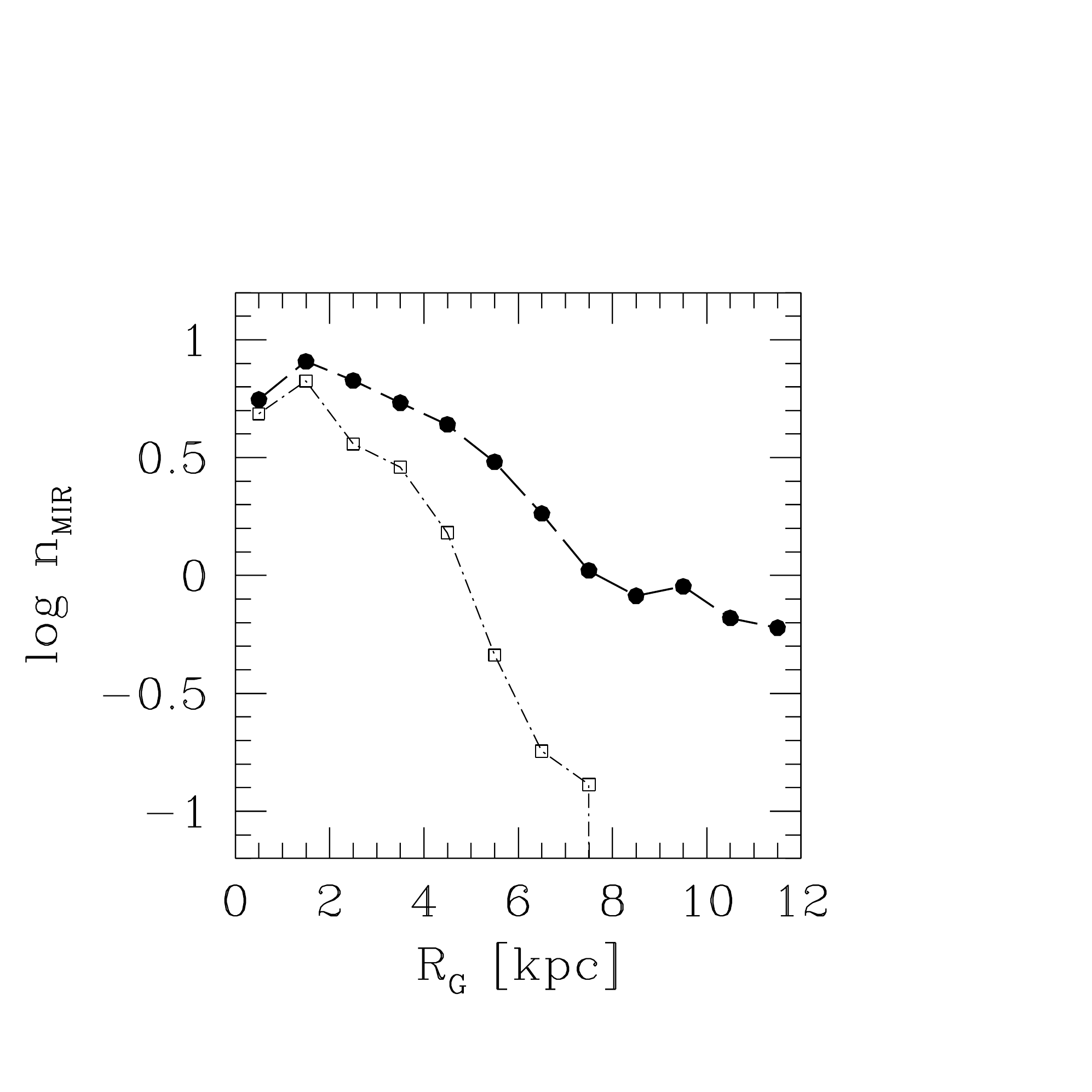}
 \caption{Number density of MIR sources, n$_{MIR}$, per kpc$^2$ (equivalent to 0.0047 deg$^2$). The
open squares and filled circles  show sources in the \citet{2011A&A...534A..96S} catalog with 24~$\mu$m
flux F$_{24}>5$~mJy and  F$_{24}<$5~mJy, respectively.  }
\label{mirs}
\end{figure}

To evaluate the level of background contamination in our sample, we plot in Figure~\ref{mirs} the radial decline of the number of MIR sources per unit area expressed in kpc$^2$. Owing to the
progressive radial decline of  star formation per unit area in M33, we expect that the number density of MIR that are true SF sites declines radially as well.  This has been shown to be the case for the
whole sample selected by \citet{2011A&A...534A..96S}, although the radial distribution flattens beyond 8~kpc. We selected 
sources with fluxes F$_{24}$>5~mJy (bright sample),  and with  5>F$_{24}$>0.2~mJy (faint sample)  and plot in Figure~\ref{mirs} the number of sources
per unit area as a function of galactocentric radius. The absence of bright sources beyond 8~kpc (open squares in Figure~\ref{mirs}) does not necessarily imply that star formation 
stops beyond this radius  since   SF sites might  be fainter at large galactocentric radius. The filled symbols connected by the heavy line in the same figure 
indicate the radial trend for  the faint sample, 
which contains the 99 sources described in Section~2 of  this paper. 
The flatness of the density distribution for fainter sources beyond 8~kpc   suggests  that this is the likely 
level of background contamination, although the baryonic surface density also flattens in the outer disk.
The  apparent weak radial decline of the faint MIR source number density in the outer disk is uncertain because of the  deconvolution procedure in the warped region (the tilted rings
partially overlap), and we considered the mean value  of the observed 
source density to estimate  the background contamination.  

Figure~\ref{mirs} shows that
the number density  of sources  per kpc$^2$ with  F$_{24}$<5~mJy and galactocentric distances between 8 and 12 kpc in  the MIR catalog varies between 0.9 and 0.6 with a mean value of  
0.75  per kpc$^2$. The 99 selected sources lie in an area of approximately 100~kpc$^2$ (this is less than what can be inferred by a simple elliptical projection  because 
there is a partial overlap of the projected areas in the warped region),  
which means that  between 60 and 90 of them can be background sources. This estimate is compatible with the higher  density inferred from 24~$\mu$m source counts in fields observed by {\it Spitzer} 
with higher sensitivity.   Although our background contamination estimate   is an upper limit  because some sources might 
be small SF sites in the outer disk of M33, the situation is clearly reversed with respect to the bright sample. 
For the faint sample the majority of the selected sources   might be background sources that are unrelated to M33, but are galaxies and quasars at high 
redshifts \citep{2004ApJS..154...80C,2007ApJ...668...45P}. This can explain the non-detection of CO lines in the selected sources.

The 24~$\mu$m fluxes for sources in the  MIR sample are between 0.4 and 2.8~mJy, while for the sample selected by \citet{2011A&A...528A.116C},  17 of the 18 SF sites
have 24~$\mu$m fluxes above 3~mJy and have on average a CO J=2-1 integrated line brightness of 1~K~km~s$^{-1}$. Although there is no exact linear scaling relation between 
the brightness of the CO lines and the 24~$\mu$m flux,  the stacked spectra and the upper limits on the CO line brightness at the location of faint MIR sources presented in this paper  
are sufficiently low  ($\le$0.1~K~km~s$^{-1}$) to suggest that  these faint sources might be unrelated to star formation in M33. We furthermore underline that none of the selected 
SF sites of \citet{2011A&A...528A.116C} was located beyond 7~kpc and
that the two detected sources in the MIR sample presented here have  rather weak MIR emission with respect to the rest of the sample, but have associated FUV and H$\alpha$ emission.

The attempt to select  truly SF sites using IRAC colors turns out to be effective for separating  MIR sources that are  SF sites  from variable stars and
local galaxies, but the lack of CO detections around 10 of the 12 observed  sources indicates that contamination from faint background sources is still present.
The CO detection in the proximity of two sources of our sample  suggests that sources with FUV luminosities  higher than $10^{36.8}$~erg~s$^{-1}$ (if placed at the distance of M33) should be
selected to find CO lines,
while  embedded sources with no UV or optical counterpart are extremely rare  because of the short duration of this SF phase \citep{2017A&A...601A.146C}.  
By discarding the 18 sources tagged as galaxies,
only 8 of the 81 sources satisfy this selection criteria, and 5 of them lie in the selected area shown in Figure~\ref{colors}. Two of these (sources s421 and s889)
are associated with cataloged GMCs
(GMC 325 and GMC 566), and two are  sources associated with the newly detected CO emission presented in this paper (sources s799 and s892). These 4 sources lie within the selected area for
IRAC colors and are within the M33 optical radius, at galactocentric distances between 7.5 and 8.1~kpc. At a similar distance lies  the only source within the selected area
with FUV luminosities  $>10^{36.8}$~erg`s$^{-1}$ that has not  been observed yet. This is s880,  a good candidate for being a SF region with associated CO emission,
although there is no H$\alpha$ emission in its proximity.  Reliable H$\alpha$  fluxes  from  compact sources in  our database have associated  luminosities   $\ge 35.8-36$~erg~s$^{-1}$ 
, and  only sources s421 and s892 of the 81 sources are clearly detected in H$\alpha$. The paucity of H$\alpha$ emission  can be explained by the IMF incompleteness in low-mass clusters, however, because the Ly-continuum
emission decreases much more rapidly than UV emission as the   stellar  mass in a young cluster decreases.  

We  underline that in the outer disk of M33, small SF sites are still present within the optical radius (8.5~kpc), they are rare and can be found by a careful analysis
of associated UV emission and IRAC colors, as confirmed by the detection of molecular lines in their close proximity presented in this paper. If our estimate of the background contamination is
correct, there might still be only a few other  MIR sources  in our sample with very faint FUV and CO luminosity that are SF sites in the outer disk of M33.  

\subsection{Evolution of the star-forming region in the southern field}

The total molecular mass recovered by mapping the southern field is about 6 10$^4$~M$_\odot$, as shown in Table~4 and discussed in Section 4.1.
Most of the molecular gas emission is in the proximity of the isolated HII region in this field, which hosts a stellar cluster whose stellar mass and age have been
estimated to be about 6000~M$_\odot$ and 8~Myr by \citet{2011A&A...534A..96S}. 
As cataloged by \citet{2013A&A...552A.140R}, the H$\alpha$ emission of MA1 in the southern field is not compact, nor is it distributed in a  pure shell 
(as in the case of RVP87, the HII region in the northern side of the outer disk),  but is of mixed morphology. This is indicative of an intermediate age
for the SF region associated with it, as confirmed by the spectral energy distribution fits by \citet{2011A&A...534A..96S}
The patchy location of CO peaks  and the high ratio of stellar  to molecular gas mass suggests that the molecular cloud is no longer contracting,  
but is affected by the stellar cluster feedback.  The velocity shift between the CO lines and the 21cm lines  at the same location in the proximity of MA1 
is larger than for the detected CO lines in the MIR sample and is consistent with this evolutionary scenario.
  
The  cluster intermediate age is supported by dust  properties: dust is still at high temperature and localized  rather than being distributed farther away, such as 
for a more evolved  shell-like  HII region.  The stellar cluster has a low infrared-to-FUV ratio,   as found  for a few other MIR sources 
located in the SF disk that have similar H$\alpha$ morphologies, 24~$\mu$m fluxes, and stellar cluster ages. None of 
these sources is coincident with massive GMCs, although one is  at the boundary of a cataloged cloud (s765) and others lie close to pixels where 
some CO emission is present. The sensitivity in the southern field map is higher than that in the all-disk survey  for the SF disk of M33, and it allows recovering weaker CO lines. In addition, cataloged GMCs have to satisfy a number of requirements concerning the spatial extent and velocity coherence
of contiguous pixels, which makes it unlikely that molecular gas distributions as patchy as  found around MA1 would be identified as one single GMC. 

The low TIR/FUV ratio, cluster age, and H$\alpha$ morphology suggest that in these regions star formation is close to its end and that stellar evolution is   
reducing the original molecular content while the cluster breaks through the gas. In a low-density environment such as the outer disk, the pressure of the 
hot gas and shock wave is unable to trigger new GMCs and episodes of star formation in the HII region proximity.
 Hence, molecular hydrogen is fading away as the  young  stellar cluster evolves.

 \section{Disk instabilities and the corotation radius}
 
\begin{figure}
\centering
\includegraphics[width=9cm]{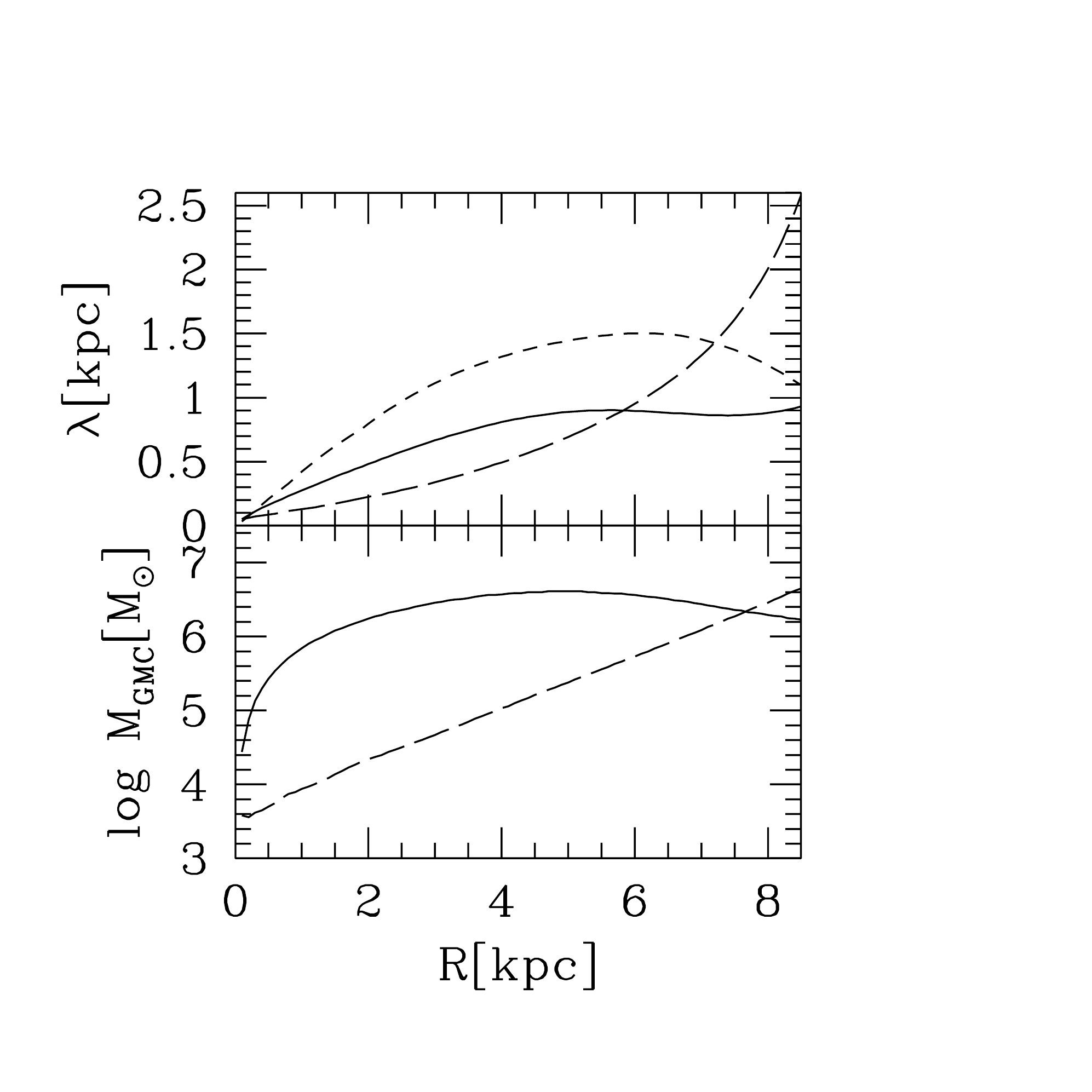}
 \caption{Unstable perturbations in the M33 disk. In the upper panel the long- and short-dashed lines show the most unstable wavelength according to Jeans criteria and to Toomre criteria, respectively, as a function of
galactocentric radius. The continuous line
indicates twice the vertical scale height for the gas. In the bottom panel  the predicted   molecular cloud masses for perturbations  with radius equal to the vertical gaseous disk scale height  
are shown with a continuous line as a function of galactocentric radius. The long-dashed line  indicates the mass corresponding to perturbations with radius equal to one-quarter the Jeans length.}
\label{unstable}
\end{figure}

 We consider the total gaseous disk of M33, made of atomic and molecular gas, which in addition to its self-gravity feels the gravity of the stellar disk.
 The  disk   is unstable according to the Toomre criterion out to about 6.5~kpc \citep{2008gady.book.....B,2003MNRAS.342..199C}. 
 The ratio of the atomic gas to stellar dispersion in M33 is about 0.5,  because the gas FWHM is radially constant and equal to about 13~km~s$^{-1}$
 and the stellar dispersion is about 25~km~s$^{-1}$, as observed in the central regions \citep{2007ApJ...669..315C,2018A&A...617A.125C}. Therefore the approximation of
 the gaseous and stellar disk stability criteria used by \citet{2003MNRAS.342..199C} and originally proposed by \citet{1994ApJ...427..759W} are reliable,
 as shown by \citet{2011MNRAS.416.1191R}. 
 Stellar kinematic maps of  galaxies of mass and morphological type similar to M33 in  the integral-field spectroscopic survey CALIFA  indicate a rather constant radial  velocity dispersion 
\citep{2017A&A...597A..48F},  and this justifies the use of the value measured in the central regions of M33 throughout the disk. 
The outermost unstable regions in the disk of M33   are between 6 and 7~kpc, and they coincide with the transition between the stellar and the 
 gas-dominated regime (see Figure 10 of \citet{2014A&A...572A..23C}). We now examine the size of the perturbations with the fastest growth rate using
 the  circular velocities, the gas, and the stellar surface density radial distributions  given by  \citet{2014A&A...572A..23C}.
The upper panel of Figure~\ref{unstable} shows the most unstable wavelength according to Jeans criteria and to Toomre criteria   as a function of
galactocentric radius. The vertical gaseous disk scale height  is also shown, and it is computed using the prescription  of  \citet{2011ApJ...737...10E} for a two-component galaxy disk, 
stars, and total gas prior to molecular cloud formation.

If perturbations are unstable, they can collapse, and as the density grows, most of the gas will become molecular; the fractional mass of the atomic envelope depends on local
conditions and on the size of  bound clouds, which is hard to determine because of the limited spatial resolution in M33.
The Jeans length gives the smallest size for  perturbations that can grow and form molecular clouds,   and this should be  smaller than the  thickness of the gaseous disk. 
 Figure~\ref{unstable} shows that this is indeed the case, and  
 clumps that later condense into molecular clouds can  grow out to about 6~kpc (considering the additional gravity 
 provided by the stellar disk) or slightly farther out if the local density is enhanced.
 We estimated the highest cloud mass as a function of galactocentric distance 
 as a product of the average local gas density times a spherical volume with radius equal  to the 
 thickness of the gaseous disk.  The bottom panel of Figure~\ref{unstable} shows this  mass and the mass corresponding to the Jeans length. Cloud masses increase radially outward in the inner disk, and  
 the maximum mass value is about 4$\times$10$^6$~M$_\odot$ at 4~kpc. Although this mass is in agreement with the highest  GMC masses recovered in M33, 
 the radial trend   is opposite to what  the data in Figure~\ref{massb} show at small galactocentric radii. 
 The time for shear to tear a gas condensation apart increases radially outwards and is longer than the free-fall time throughout the SF disk, which means that shear cannot be
 responsible for the radial decrease of the molecular cloud mass  presented in Section~2 either. 
 
 As shown in Figure~\ref{unstable},  the most unstable wavelength increases radially outward, and this justifies 
 the radial decrease in number density of molecular clouds from the center to the outer disk.  The most unstable Toomre length is   indicative of the separation between filaments 
 that form in the unstable disk. Therefore the disk instability analysis correctly predicts the observed drop in number density of 
 molecular clouds in the stable outer disk, but  it cannot explain the extra growth of individual perturbations in the inner disk where the mean GMC mass is observed to increase toward the center.  
 In the next subsection we examine an additional mechanism, the rotation of the main arm pattern, which can play a role in driving the radial decrease in molecular complex mass in the inner disk. 
This decrease is not fast, and it is indicative that this mechanism   will pile up only a few clouds to provide the extra growth in mass. Only close to the galaxy center is the difference between the observed  
and the predicted mass (bottom panel of Figure~\ref{unstable}) more evident.  However,  here the gas vertical scale height is  likely underestimated by the model. Furthermore, in the central 
regions the ISM is highly turbulent, as  the shape of the probability distribution  function shows \citep{2018A&A...617A.125C},   and the high rate of star formation likely triggers the formation of new clouds 
and enhances their  agglomeration into larger complexes by  frequent episodes of gas compression.

\subsection{Corotation radius}

\begin{figure} 
\includegraphics[width=9cm]{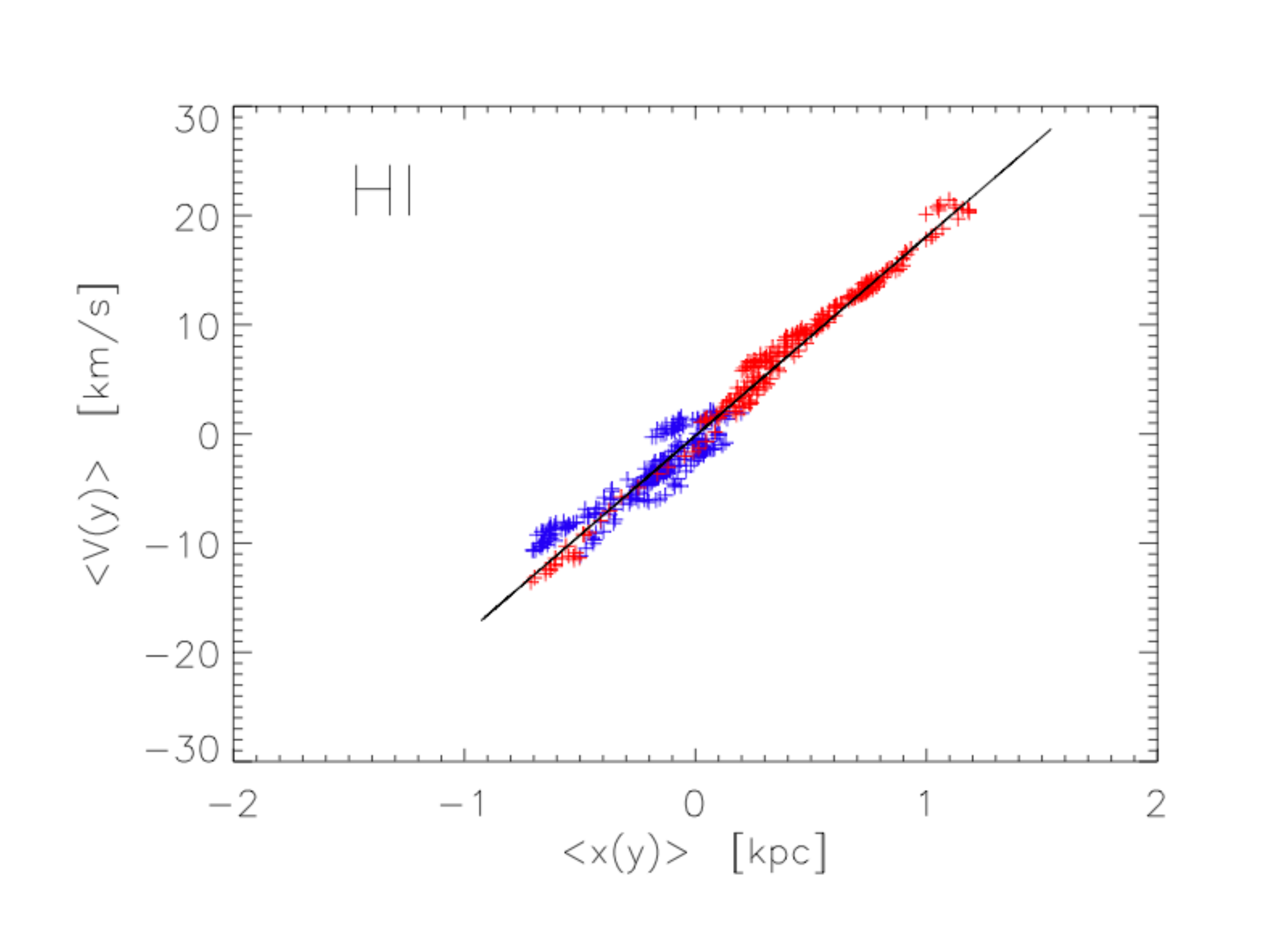}
\caption{Intensity-weighted mean line-of-sight HI velocity defined by Eq. \ref{Vy}
vs. the
intensity-weighted mean position defined by Eq. \ref{xy}.
Each point represents a different horizontal strip in the moment images,
at distance $y$ from the major axis. In red we show data from the far side to the east of the major axis,
and in blue we plot data from the near side, to the west of major axis.
The solid line is the fitted regression, whose slope
gives $\Omega_P\times sin~i$. }
\label{HIpattern}
\end{figure}

\begin{figure}
\includegraphics[width=9cm]{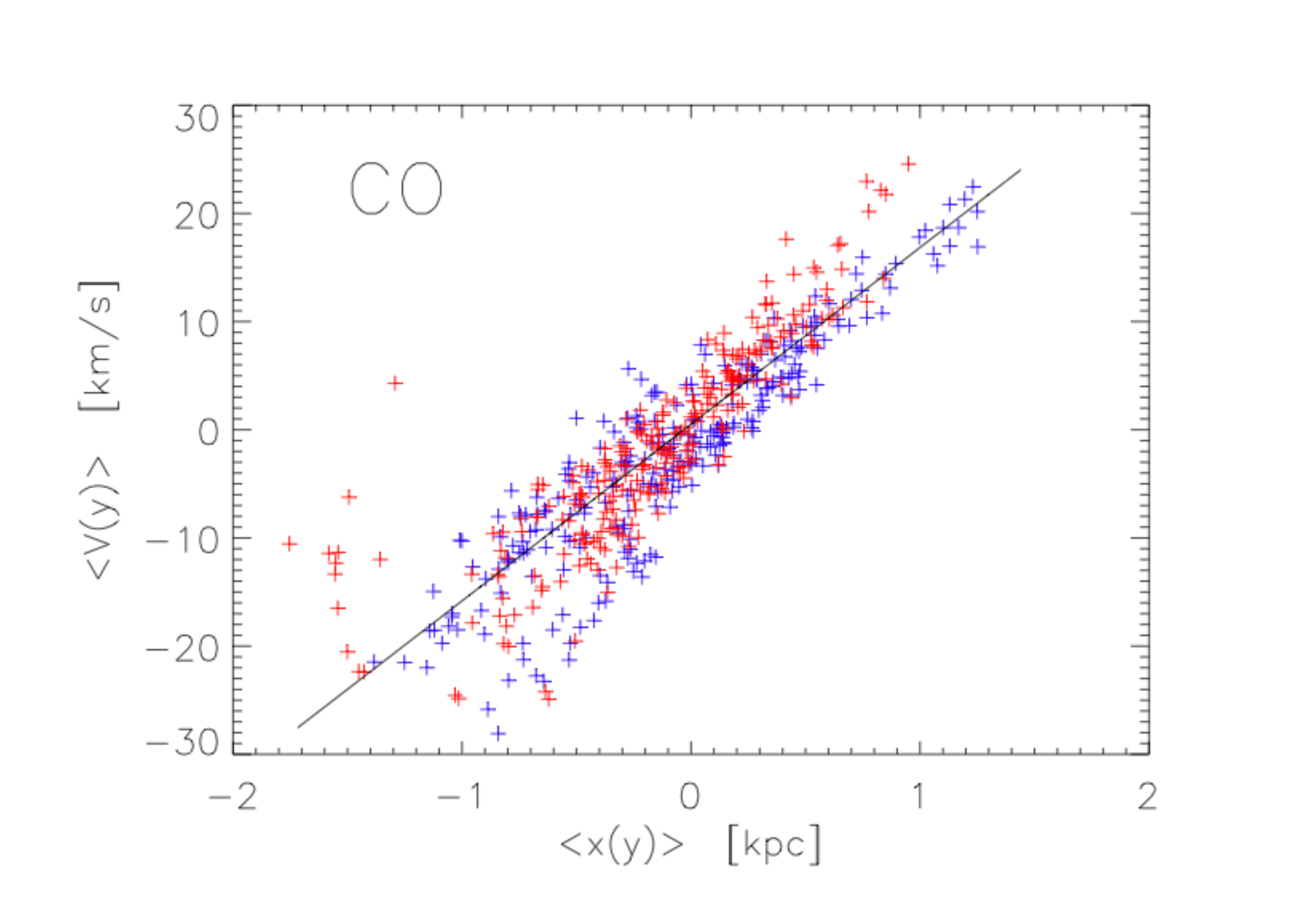}
\caption{Same as in\ Figure \ref{HIpattern}, but for the CO(2-1) moment images.}
\label{COpattern}
\end{figure}

In order to derive the speed of the spiral pattern in the disk of M33,
we adopted the kinematical method developed by \citet[TW]{1984ApJ...282L...5T}.
The method was originally devised to provide a model-independent means of measuring 
the pattern speed of bars in SB0 galaxies. It has subsequently been generalized to
non-rigid patterns, that is, with a radially varying pattern speed, 
by \citet{2006MNRAS.366L..17M}, and was used to estimate the pattern speed in nearby
galaxies using \hi\ or CO emission as mass tracer by \citet{ 2004ApJ...607..285Z}, for instance.
It is assumed that the disk is flat and thin with negligible vertical motions;
it is also assumed that the intensity in the adopted tracer, as representative of the 
surface density of a mass component, obeys a continuity equation along the orbit
around the galactic center, 
meaning that there is no source or sink of the tracer
\footnote{
In this respect, the components of the ISM do not constitute an ideal tracer, 
but we can still assume that for the surface density of the
component chosen, the relative variations in a parcel of gas are not 
dramatic within the single orbit. 
}.
Both the inclination to the line of sight $i$ and the position angle $PA$ 
are assumed to be known. 
Under these hypotheses, the derivation of the pattern speed is quite straightforward.
First we produce zeroth- and first-moment images from the 21cm and CO(2-1) cubes and 
rotate them to align the major axis horizontally ($x$-axis, positive to the receding SW side).
After subtracting the systemic velocity $V_{Hel}=-180$~km~s$^{-1}$, 
for each strip of pixels parallel to the major axis at distance $y$ from it, we 
evaluate the intensity-weighted line-of-sight mean velocity

\begin{equation}
\langle V(y) \rangle = \frac {\int_{-\infty}^{+\infty} I(x,y)~v_{LOS}(x,y)~dx}{\int_{-\infty}^{+\infty} I(x,y)~dx}
\label{Vy}
,\end{equation}

\noindent and similarly, the intensity-weighted mean x-position of the tracer
\begin{equation}
\langle x(y) \rangle = \frac {\int_{-\infty}^{+\infty} I(x,y)~x~dx}{\int_{-\infty}^{+\infty} I(x,y)~dx}
\label{xy}
.\end{equation}

\noindent According to the TW method, the angular speed of the pattern is then given by
\begin{equation}
\Omega_P=\frac{1}{sin~i} \frac{\langle V(y) \rangle}{\langle x(y) \rangle}
.\end{equation}

For each strip, at each of the sampled $y$, we will have a pair of 
$\langle V(y) \rangle$,
$\langle x(y) \rangle$ values that are linearly related in case of a rigid pattern.
The slope of the relation $\langle V(y) \rangle$ versus $\langle x(y) \rangle$
is $\Omega_P\times sin~i$.

The results are shown in Figures \ref{HIpattern} and \ref{COpattern}. 
The point distribution is clearly linear and therefore indicative
of a well-defined rigid trailing spiral pattern;
in addition, the results from the two tracers are remarkably consistent. 
We obtain the tightest regressions, 
and best agreement between \hi\ and CO results, by adopting a 
$ PA=23^\circ\pm 2^\circ $
, a value quite close to the one derived from the optical images
(e.g., $22.7^\circ$ in HyperLeda  or  $23^\circ$ in \citet{1991rc3..book.....D}).
From the HI we obtain
$\Omega_P\times sin~i = 18.15 \pm 0.15~{\rm km~s}^{-1} {\rm kpc}^{-1}$, and from the CO(2-1)
we get $18.05 \pm 0.35$ in the same units.
The final estimate is then $18.13 \pm 0.14~{\rm km~s}^{-1} {\rm kpc}^{-1}$. 

Using the rotation curve and the (inner) disk inclination ($i$=54$^\circ$) in
\citet{2014A&A...572A..23C}, we finally obtain a corotation radius
$R_{cor}=4.7 \pm 0.3~{\rm kpc}$. The assigned uncertainty includes those on
inclination, rotation curve, and $\Omega_P\times sin~i$.
Assuming the 21cm second moment measures the sound speed (including turbulence),
we can locate the sonic point, 
where the velocity difference between pattern and rotating gas 
equals the sound speed, at a radius of $3.9~{\rm kpc}$.
We also performed the preceding analysis separately for the two halves of the galaxy,
NE and SW of the minor axis. In this case, the corotation is more external in the NE,
$5.4~{\rm kpc}$, than SW, $4.5~{\rm kpc}$; we find no significant difference in
the mean velocity dispersion.

 We summarize the importance of the corotation radius as follows:  the lack of molecular clouds  in the  outer disk as well as the mass distribution of GMCs across the SF disk of M33 is 
 consistent with the mass predicted by the growth of  individual condensations in the gaseous unstable disk, except in the innermost  regions, where  the observed mean GMC mass increases toward the center. 
When we account for the fast disk rotation with respect to the arm pattern, this discrepancy is alleviated because molecular clouds can experience an extra growth as the arms collect smaller clouds into larger 
self-shielded complexes inside the corotation radius. The location of the corotation radius is consistent with the galactocentric radius inside which the  extra growth of GMC mass is observed.

\section{Summary and conclusions}

We have presented  the results of deep observations of CO lines at selected locations in the outer disk of M33 to complement the results of existing surveys of molecular clouds
in the SF disk and examine possible triggers of  molecular cloud mass growth  across the disk of the closest blue disk galaxy. We detected  CO J=1-0 and J=2-1 lines  near the position of
two MIR sources in the outer disk, at  about 8~kpc from the center, which have 24~$\mu$m fluxes of only 0.5~mJy and  faint UV and optical counterparts.  
The  CO lines are very narrow, with main-beam peak temperatures 
of about 0.05-0.1~K. At the location of another ten selected MIR with 24~$\mu$m fluxes between 0.4 and 2.8~mJy and galactocentric distances between 
7.7 and 10.4~kpc,  we have only upper limits on the CO line emission that are considerably below the line brightness detected in the proximity
of the other two MIR sources.

We also detect CO J=2-1 lines in the proximity of MA1, one of the outermost  HII regions in the south, at about 7.4~kpc from the center. We mapped  the CO J=2-1 line emission in an area of about  
5~arcmin$^2$  and detect CO lines that imply a total molecular mass of about 6$\times10^4$~M$_\odot$. The most prominent line is emitted  close to the H$\alpha$ brightness peak where 
most of the dust  emission is also located. The other CO lines are detected at the boundary 
of the  HII regions: here the gas is likely compressed through shock fronts that develop as  the HII region expands, and some cold gas, molecular and atomic, can be found.

It is conceivable that the paucity of CO lines in the outer disk of M33 is enhanced by the lower metal abundance, although the metallicity gradient in the bright disk is not very steep. However,
the analysis in this paper has shown that a change in  the disk dynamical conditions is likely at the heart of the detected variations of molecular cloud growth across M33. We summarize our main conclusions below.

\begin{itemize}

\item
The speed of the spiral arm pattern in the inner disk is determined through a kinematical method applied to both HI and CO data sets. The atomic and molecular gas
give consistent results: the arms of M33 rotate at
about 22.4~km~s$^{-1}$~kpc$^{-1}$. This speed is slower than the disk rotation inside the corotation radius, which for M33 is 4.7~kpc. Given the HI velocity dispersion, we locate the
sonic point at 3.9~kpc. This implies that inside 3.9~kpc,  the arm pattern   is able to collect gas condensations into larger clouds and trigger star formation through shocks. 

\item
The Toomre criterion predicts that the disk is stable beyond 6.5~kpc if gravity from stars and gas is considered. Inside 6.5~kpc, filaments and gas perturbation can grow. The  
most unstable wavelength, larger than the Jeans length,   increases from the center of the galaxy out to about 6~kpc. As the disk vertical thickness
is smaller than the  most unstable wavelength, the disk vertical scale height sets the maximum mass of molecular complexes, which is rather constant and equal to the
measured maximum mass of observed GMCs.   In the central region,  only low-mass GMCs should be found, but this is not the case.

\item
The mean molecular mass of cataloged complexes in M33 is instead higher at the galaxy center and decreases radially by a factor 2 out to 4~kpc, then it flattens out to about 6.5~kpc 
and drops beyond this radius.  The observed trend for R$<4$~kpc can be reconciled with the expected decrease of GMC masses toward the center  predicted by disk dynamics when the slow
rotation of the arm pattern inside corotation is considered. Arms can collect molecular clouds into larger complexes especially where the disk motion relative to them  is supersonic. 

\item
The population of faint MIR sources beyond the SF disk in the sky area where the warped outer disk of M33 is found is consistent with the density of background source counts at
24$\mu$m. This explains the paucity of CO lines detected at the location of selected MIR sources and confirms that SF sites beyond the disk unstable region are rare. 
Stacking of CO J=1-0 and J=2-1 spectra for undetected MIR sources has not unveiled any fainter CO emission.
However, narrow CO lines have been detected at the location of two MIR sources associated with  FUV emission  even in the outer disk,  and molecular cloud masses have been
estimated.

\item
The patchy distribution of detected CO lines in the outer disk region hosting MA1, the local weak MIR-to-FUV ratio, and the morphology of the ionized gas suggest that in MA1 star formation
is close to its end, as the HII region expands and dissipates the cold gas. This is is consistent with the
age and mass estimate of the YSC hosted by MA1.  
  
\end{itemize}
 
Our results imply that  internal disk dynamics can then play a crucial role in  
 regulating the growth of  gas perturbations that later collapse and form stars across the disk of M33. Star formation sites in the outer disk  are not widespread at the present cosmic time.
The population of intermediate-mass stars in the far outer disk \citep{2011A&A...533A..91G}, where no bright FUV knots are found,
is likely the result of a past burst of star formation  triggered by gas or satellite accretion \citep{2008A&A...487..161G,2018MNRAS.480.4455M}. Stellar migration from the inner disk 
 cannot affect regions that are in the far outer disk in less than 1~Gyr \citep{2016A&A...588A..91M}.
 CO is found at the location of  few MIR sources with FUV emission in the outer disk where the azimuthally averaged  HI column density is still above 1~M$\odot$ \citep{2014A&A...572A..23C}.
 This is consistent with the location of  young stellar clusters associated with FUV knots in dwarf irregular galaxies \citep{2016AJ....151..136H}, where the  larger
 distance and lower metallicity prevent deep local CO searches.

\begin{acknowledgements}
We would like to thank the anonymous referee for their comments and careful reading of the original manuscript. E.C. thanks Bruce Elmegreen for  stimulating discussion on topics related to this paper.
 We acknowledge funding from the INAF PRIN-SKA 2017 program 1.05.01.88.04.
\end{acknowledgements}

\end{document}